\documentclass[a4paper,12pt]{article}
\usepackage{graphicx,amssymb,bm,latexsym,color,epsf}
\usepackage{amsmath}
\usepackage{appendix}
\usepackage{comment}
\usepackage{xcolor}
\usepackage[normalem]{ulem}
\usepackage{hyperref}
\usepackage{mathrsfs}
\usepackage[utf8]{inputenc}

\usepackage[
backend=biber,
style=numeric-comp,
sorting=none,
maxbibnames=99,
giveninits=true,
isbn=false
]{biblatex}

\hypersetup{
colorlinks=true,
citecolor=blue,
citebordercolor=red,
linktoc=all,
linkcolor=blue,
urlcolor=blue
}

\makeatletter
\@addtoreset{equation}{section}

\makeatother
\newcommand{\bea}{\begin{eqnarray}}
\newcommand{\eea}{\end{eqnarray}}
\newcommand{\vs}[1]{\vspace{#1 mm}}
\newcommand{\hs}[1]{\hspace{#1 mm}}
\renewcommand{\a}{\alpha}
\renewcommand{\b}{\beta}
\newcommand{\G}{\Gamma}
\renewcommand{\d}{\delta}

\newcommand{\s}{\sigma}
\renewcommand{\t}{\theta}

\newcommand{\la}{\lambda}
\newcommand{\pa}{\partial}

\newcommand{\nn}{\nonumber\\}
\newcommand{\p}{\partial}

\newcommand{\br}{\bar R}
\newcommand{\bg}{\bar g}

\newcommand{\bnabla}{\bar\nabla}

\newcommand{\Tr}{{\rm Tr}}
\newcommand{\Det}{{\rm Det}}

\newcommand{\Dsl}{D \kern-0.65em /\,}
\newcommand{\dsl}{\partial \kern-0.55em /\,}
\newcommand{\Fsl}{F \kern-0.65em /\,}
\newcommand{\lich}{{\Delta_L}}

\newcommand{\pmat}[1]{\begin{pmatrix}#1\end{pmatrix}}

\allowdisplaybreaks
\begin{document}

\begin{flushright}
\today
\end{flushright}
\renewcommand{\thefootnote}{\fnsymbol{footnote}}
\newbox{\ORCIDicon}
\sbox{\ORCIDicon}{\large
                  \includegraphics[width=0.8em]{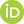}}

\begin{center}
{\Large\bf Higgs scalar potential coupled to gravity in the exponential parametrization in arbitrary gauge
}
\vs{5}

{\large
Nobuyoshi Ohta\href{https://orcid.org/0000-0002-8265-8636}{\usebox{\ORCIDicon}},$^{1,2,}$\footnote{e-mail address: ohtan@ncu.edu.tw}
and Masatoshi Yamada{}\href{https://orcid.org/0000-0002-1013-8631}{\usebox{\ORCIDicon}}$^{3,}$\footnote{e-mail address: m.yamada@thphys.uni-heidelberg.de}
} \\
\vs{5}

$^1${\em Department of Physics, National Central University, Zhongli, Taoyuan 320317, Taiwan}
\vs{3}

${}^2${\em Research Institute for Science and Technology,
Kindai University, Higashi-Osaka, Osaka 577-8502, Japan}
\vs{3}

$^3${\em Institut f\"ur Theoretische Physik, Universit\"at Heidelberg, Heidelberg, Germany}

\vs{5}
{\bf Abstract}
\end{center}

We study the parametrization and gauge dependences in the Higgs field coupled to gravity in the context of asymptotic safety.
We use the exponential parametrization to derive the fixed points for the cosmological constant, Planck mass, Higgs mass and its coupling, keeping arbitrary gauge parameters $\a$ and $\b$, and compare the results with the linear split.
We find that the beta functions for the Higgs potential are expressed in terms of redefined Planck mass such that the apparent gauge dependence is absent.
Only the trace mode of the gravity fluctuations couples to the Higgs potential and it tends to decouple in the large $\b$ limit, but the anomalous dimension becomes large, invalidating the local potential approximation.
This gives the limitation of the exponential parametrization.
There are also singularities for some values of the gauge parameters but well away from these, we find rather stable fixed points and critical exponents.
We thus find that there are regions for the gauge parameters to give stable fixed points and critical exponents against the change of gauge parameters.
The Higgs coupling is confirmed to be irrelevant for the reasonable choice of gauge parameters.


\renewcommand{\thefootnote}{\arabic{footnote}}
\setcounter{footnote}{0}

\section{Introduction}

The functional renormalization group~\cite{Wilson:1973jj,Wegner:1972ih,Polchinski:1983gv,Wetterich:1992yh,Morris:1993qb,Ellwanger:1993mw}
is a powerful method to tackle non-perturbative phenomena in quantum field theory. For reviews,
see~\cite{Morris:1998da,Berges:2000ew,Aoki:2000wm,Bagnuls:2000ae,Polonyi:2001se,Pawlowski:2005xe,Gies:2006wv,Delamotte:2007pf, Rosten:2010vm,Braun:2011pp,Dupuis:2020fhh}.
Its application to gauge theories has elucidated the non-perturbative nature of gauge
theories~\cite{Reuter:1993kw,Reuter:1996ub,Gies:2002af,Gies:2003ic,Gies:2006wv,Braun:2014ata,Mitter:2014wpa,Cyrol:2016tym,Cyrol:2017ewj,Corell:2018yil}.
In particular, the functional renormalzation group has contributed towards asymptotically safe quantum gravity~\cite{Hawking:1979ig,Reuter:1996cp,Souma:1999at}
which is formulated as a non-perturbative quantum field theory. Its recent developments are summarized in Refs.~\cite{Niedermaier:2006wt,Niedermaier:2006ns,Percacci:2007sz,Reuter:2012id,Codello:2008vh,Eichhorn:2017egq,Percacci:2017fkn,Eichhorn:2018yfc,Reuter:2019byg,Wetterich:2019qzx,Bonanno:2020bil,Reichert:2020mja,Pawlowski:2020qer}.
A central object in the functional renormalization group is the effective average action $\Gamma_k$ and its second-order functional derivative $\Gamma_k^{(2)}$ whose inverse form corresponds to the full propagator, so that in general the gauge fixing is required for the gauge field propagator in addition to the ultraviolet (UV) regularization.

Physical quantities should depend on neither gauge fixing parameters nor regularization schemes;
however, such an independence may be easily destroyed by the gauge fixing and approximations in the quantum theory.
Therefore, we need to carefully handle their dependence when we apply the functional renormalization group especially to gauge systems. 
Good gauge choice may be identified by allowing for a family of gauge and identifying stationary points
in the parameter space. This is the principle of minimum sensitivity advocated in \cite{Stevenson:1981vj,Ball:1994ji,Gies:2015tca}.

Recently the effective potential for the Higgs field was calculated in various works~\cite{Percacci:2003jz,Narain:2009fy,Narain:2009gb,Percacci:2015wwa,Oda:2015sma,Labus:2015ska,Hamada:2017rvn,Eichhorn:2017als,Pawlowski:2018ixd,Wetterich:2019zdo}.
It is an important quantity to understand the electroweak symmetry breaking in the Standard Model of particle physics.
The fixed points of the quantum effective potential for the Higgs field was studied by calculating
beta functions of the couplings in the theory, and it has been shown that quartic self-interaction of
the Higgs scalar field is an irrelevant coupling at the asymptotically safe UV fixed point of quantum gravity.
This has an important prediction to the ratio of the masses of the Higgs boson and top quark~\cite{Shaposhnikov:2009pv,Eichhorn:2017eht,Eichhorn:2017ylw,Pawlowski:2018ixd,Eichhorn:2017als,Eichhorn:2018whv,Alkofer:2020vtb}.
Moreover, the understanding of the energy scaling of the Higgs mass parameter is essential towards the gauge hierarchy problem. It was pointed out in Ref.~\cite{Wetterich:2016uxm}
that quantum-gravity fluctuations make the Higgs mass parameter irrelevant and thus could play a key role for solving the gauge hierarchy problem.
The determination whether the number of the relevant coupling constants is finite and how many are there are also crucial questions for obtaining predictable theory of quantum gravity. By including higher curvature terms~\cite{Codello:2006in,Benedetti:2009rx,Benedetti:2009gn,Ohta:2013uca,Ohta:2015zwa}, these problems have been studied in pure gravity~\cite{Benedetti:2013jk,Falls:2013bv,Falls:2014tra,Falls:2017lst,Falls:2018ylp,Falls:2020qhj,Kluth:2020bdv}.
Extensions to matter system including standard model of particle physics and beyond are considered~\cite{Eichhorn:2017als,Reichert:2019car,Hamada:2020vnf,Kowalska:2020zve,Eichhorn:2020kca,Eichhorn:2020sbo,Eichhorn:2021tsx}. See also Refs.~\cite{Eichhorn:2017muy,DeBrito:2019rrh,Kowalska:2020gie} as applications of the asymptotic safety scenario for gravity-matter systems.

Although these works represent important progress, it is also known to suffer from the gauge dependence
in the gravity contribution~\cite{Falkenberg:1996bq,Souma:2000vs,Barra:2019rhz}. The question arises then how much the obtained results are reliable
in the physics of spontaneous breaking of the symmetry.
We would like to study this problem by keeping the gauge parameters, and check how much the results depend on them.

Another possible problem is that it is found that there appears some unphysical poles in the cosmological
constant in the beta functions for the gravity couplings if one uses the linear parametrization
of the metric
\bea
g_{\mu\nu}=\bg_{\mu\nu}+h_{\mu\nu}.
\label{lin}
\eea
Due to this, it appears the results are unstable close to the singularity.
It is possible that the sign of the critical exponents changes near the singularity, thus drastically changes the nature of the fixed points~\cite{Eichhorn:2017als}.
This might be a signal of phase transition, but this cannot be treated in the polynomial truncation. In the present setting, it would be better to consider the region away from the singularity.
In this respect, it is known that such singularity in the cosmological constant does not appear
if one uses the exponential parametrization for the gravitational fluctuation~\cite{Percacci:2015wwa}:
\bea
g_{\mu\nu} = \bar{g}_{\mu\la}(e^h)^\la{}_\nu = \bg_{\mu\nu}+h_{\mu\nu}+ \frac{1}{2} h_{\mu\la}h^\la_\nu + O(h^3).
\label{exp}
\eea
The parametrization was first used in \cite{Kawai:1992np}, applied to $D=4$ gravity in asymptotic safety~\cite{Ohta:2015efa,Ohta:2015fcu,Falls:2016msz}
and later generalized to more general parametrization to study various parametrization dependence~\cite{Gies:2015tca,Ohta:2016npm,Ohta:2016jvw,deBrito:2018jxt,Goncalves:2017jxq}.
Here we choose this exponential parametrization in order to avoid the unphysical singularity and check
the consistency with the earlier results.
This parametrization has also the advantage that the resulting beta function does not depend
on the gauge choice~\cite{Nink:2014yya,Falls:2015qga}.
However we will find that this parametrization also has some problem in the local potential approximation.

In this work, using the Wetterich equation~\cite{Wetterich:1992yh}, we study the parametrization and gauge dependences of the fixed point structure and the critical exponents in the Higgs-gravity system
by comparing the results with the exponential~\eqref{exp} and linear~\eqref{lin} parametrizations with arbitrary gauge parameters, and check if we can find reasonable results despite these dependences.
It turns out that if we use a redefined Planck mass, the beta functions for the scalar potential in the exponential parametrization become independent of the gauge parameters,
and we can discuss the fixed points without apparent gauge dependence.
This is an advantage of the exponential parametrization, but the price is that the anomalous dimension of the Higgs scalar
becomes large in general.
However we have found that its effect is actually not so big owing to the suppression factor like $1/4$.
We have found that there is a reasonable range of gauge parameters which
give stable results against change of the parameters.
Relying on the principle of minimal sensitivity, we conclude that the coupling of the self-interaction of the Higgs field is irrelevant.

This paper is organized as follows:
In Section~\ref{sec: action and hessians}, we introduce the effective action for the Higgs-gravity system with two arbitrary gauge parameters.
The two-point functions, i.e. the Hessians, are also derived.
The explicit forms of the beta functions for the Higgs-gravity system are derived with the full dependence on the gauge parameters in Section~\ref{sec: Flow equations}.
The anomalous dimension for the Higgs scalar is also given.
In Section~\ref{sec: Linear versus exponential parametrization}, we discuss the fixed points and critical exponents in the exponential parametrization for various choices of the gauge parameters, and comparae the results in the linear parametrization.
In Section~\ref{sec: gauge dependence}, we further study the gauge parameter dependence of the fixed points and anomalous dimensions.
Section~\ref{sec: conclusions} is devoted to the summary of our results and conclusions.
Technical details of the calculations and definitions are relegated to Appendices.
In Appendix~\ref{york}, we give the York decomposition and discuss the Jacobian which arises in changing the variables.
In Appendix~\ref{lich}, we give the definition and properties of the Lichnerowicz Laplacians.
Some technicality in the variation of the action is summarized in Appendix~\ref{app:sec:variation}.
The flow equations and anomalous dimension of the Higgs field are calculated in Appendix~\ref{app: sec: flow equation}.

\section{Action and Hessians}
\label{sec: action and hessians}

\subsection{Gauge fixed action for the Higgs coupled to gravity}
\label{sec: Higgs coupled to gravity}

We would like to calculate the beta functions for Higgs couplings, Newton constant and cosmological constant
with arbitrary gauge fixing parameters, in order to check gauge dependence explicitly.
The effective action we consider is given by
\bea
\G_k=\G^{\rm gravity}+ \G^{\rm Higgs}.
\label{eq: effective action in this work}
\eea
The gravity part $\G^{\rm gravity}$ consists of the Einstein-Hilbert term, gauge fixing and Faddeev-Popov ghost
terms given by
\bea
\G^{\rm gravity} &=& - Z_N \int d^4 x \sqrt{g} R + \G_{\rm GF}+\G_{\rm gh}, \nn
\G_{\rm GF} &=& \frac{1}{2\a} \int d^4 x \sqrt{\bg} \bg^{\mu\nu}f_\mu f_\nu, \\
\G_{\rm gh} &=& \int d^d x\sqrt{\bg}\,\bar C_\mu\left( \bar\nabla^2\d_\nu^\mu +\frac{1-\b}{2}
\bnabla^\mu \bnabla_\nu +\bar R_\nu^\mu \right) C^\nu,
\nonumber
\eea
where $Z_N=1/(16\pi G)$, $\bar{g}={\rm det}(\bar{g}_{\mu\nu})$, $\a$ and $\b$ are (dimensionful and dimensionless) gauge fixing parameters, and the gauge fixing function $f_\mu$ is
\bea
f_\mu &=& \bnabla^\nu h_{\mu\nu} -\frac{1+\b}{4} \bnabla_\mu h.
\label{gff}
\eea
Here the full metric is split into background $\bg_{\mu\nu}$ and fluctuation $h_{\mu\nu}$
by the exponential parametrization~\eqref{exp} and $h=\bar g^{\mu\nu}h_{\mu\nu} = h^\nu{}_\nu$ is the trace mode in the fluctuation field.
The covariant derivative $\nabla_{\mu}$ is a general relativity metric covariant derivative constructed
using Levi-Civita connection, and barred one is constructed with the background metric.
We keep both the gauge parameters $\a$ and $\b$ arbitrary in the following analysis in order to study the gauge dependence
of the results.

The Higgs field is a component of the doublet field coupled to the $SU(2)_L$ and $U(1)_Y$ gauge fields
as well as to quarks and leptons. The contributions from those fields to the beta function are smaller than
the ones from the graviton near the UV fixed point, and we can safely restrict the discussions to a single
scalar field with $\mathbb Z_2$ symmetry. The action for the Higgs field is
\bea
\G^{\rm Higgs} = \int d^4 x \sqrt{g}\left[ \frac{Z_\phi}{2}g^{\mu\nu}\pa_\mu\phi\pa_\nu\phi +U(\rho)\right],
\eea
where the effective potential is assumed to depend only on the invariant $\rho=\phi^2/2$.

To derive the flow equations in the system \eqref{eq: effective action in this work}, we use the Wetterich equation~\cite{Wetterich:1992yh} whose form reads
\begin{align}
    \p_t \Gamma_k =\frac{1}{2}\Tr \left[ \left(\Gamma_k^{(2)} + \mathcal R_k \right)^{-1}\p_t\mathcal R_k \right],
\end{align}
with $t=\log k$ the dimensionless scale.
Here, $\mathcal R_k$ is a regulator function and $\Gamma_k^{(2)}$ is the full two-point function, i.e. the so-called Hessian.
In the next section, we show the Hessians for the effective action \eqref{eq: effective action in this work}.

\subsection{Hessians}
\label{sec: Hessian}

We consider the theory on the Einstein space in which the Ricci tensor is given by
\bea
\br_{\mu\nu}=\frac{\br}{4} \bg_{\mu\nu}.
\label{es}
\eea
Using the York decomposition described in Appendix~\ref{york}, we can read off
the Hessian for the gravity from, for example \cite{Ohta:2015fcu}:\footnote{There are typos in the last terms
in Eqs.~(2.8) and (2.9) in Ref.~\cite{Ohta:2015fcu}; the factor should be 1 instead of 2.}
\bea
I^{\rm gravity(2)} &=& Z_N \left[~ \frac{1}{4} h^{TT}_{\mu\nu} \left(\lich_2 -\frac{\br}{2} \right)h^{TT,\,\mu\nu}
-\frac{3}{32} \s \lich_0^2 \left(\lich_0-\frac{\br}{3} \right)\s \right.
\nn &&
\left. ~~~~~ -\frac{3}{16} h \lich_0 \left(\lich_0-\frac{\br}{3}\right) \s
-\frac{3}{32} h \left(\lich_0+\frac{\br}{3}\right) h \right].
\label{Hessg}
\eea
Here, $\Delta_{Li}$ are the Lichnerowicz Laplacians defined in Appendix~\ref{lich}, and hereafter we drop the bars on the covariant derivatives and Lichnerowicz Laplacians.
We will rescale the graviton fields as $h_{\mu\nu} \to Z_N^{-1/2}\, h_{\mu\nu}$ so that
we can get rid of this factor from the Hessian, though it will appear in other terms.
Note that in the exponential parametrization, the gauge mode $\xi_\mu$ completely decouples from the
gauge invariant action.
The gauge fixing term then reduces to
\begin{align}
I_{\rm GF} &= \frac{1}{2\tilde\a}\int d^4 x \Bigg[\xi_\mu \left(\lich_1-\frac{\br}{2}\right)^2 \xi^\mu
+\frac{9}{16}\s \lich_0\left(\lich_0-\frac{\br}{3}\right)^2 \s
\nn
&\qquad\qquad\qquad\qquad\qquad
+\frac{3\b}{8}\s\lich_0 \left(\lich_0-\frac{\br}{3}\right) h +\frac{\b^2}{16}h \lich_0 h \Bigg],
\label{gf}
\end{align}
where $\tilde\a = Z_N\alpha$ is the dimensionless gauge fixing parameter.

The Faddeev-Popov ghost is decomposed as
\bea
C^\mu = C^{T\mu}+\nabla^\mu\frac{1}{\sqrt{\lich_0}} C^L,
\label{eq: decomposition of ghost field}
\eea
with $\nabla_\mu C^{T\mu}=0$, and the same for $\bar C_\mu$.
The ghost action reduces to
\bea
\G_{\rm gh} = \int d^4 x\sqrt{\bg}\left[ -\bar C^T_\mu\left(\lich_1-\frac{\br}{2}\right)C^{T\mu}
-\frac{3-\b}{2}\bar C^L\left(\lich_0-\frac{\br}{3-\b}\right) C^L\right].
\label{ghostkin}
\eea

Since we are interested in the corrections to the Higgs potential, we neglect those terms
where the derivatives are acting on the (background) scalar fields.
The Higgs contribution is then
\bea
I_{\rm Higgs}^{(2)} = \varphi\left[ \frac{Z_\phi}{2} \lich_0 + \frac{1}{2} (U'+2 U'' \rho )\right]\Bigg|_{\phi=\bar\phi} \varphi
+ \frac{1}{2} Z_N^{-1/2} h U' \varphi + \frac{1}{8}U h^2,
\eea
where the prime is the derivative with respect to $\rho$, and $\bar\phi$ and $\varphi$ are background and fluctuation fields of $\phi$, respectively.
Note that only the trace mode of the gravity fluctuation couples to the Higgs potential in the exponential parametrization.

The Hessian for transverse-traceless (TT) mode is given by
\bea
\frac{1}{2} h^{TT,\,\mu\nu} \G^{TT}_{\mu\nu,\,\a\b} h^{TT,\,\a\b}, \qquad
\G^{TT}_{\mu\nu,\,\a\b}=\frac{1}{2}\left(\lich_2-\frac{\br}{2}\right) E_{\mu\nu\alpha\beta} ,
\label{eq: Hessian for TT mode}
\eea
with the unity matrix $E_{\mu\nu\alpha\beta}=\frac{1}{2}(\bg_{\mu\a}\bg_{\nu\b}+\bg_{\mu\b}\bg_{\nu\a})$.
The Hessian for spin-1 transverse mode comes only from the gauge fixing term~\eqref{gf} and is given by
\bea
\frac12 \xi^\mu \G^{\xi\xi}_{\mu,\nu} \xi^\nu, \qquad
\G^{\xi\xi}_{\mu,\nu}=\left(\lich_1-\frac{\br}{2}\right)^2\bg_{\mu\nu},
\label{eq: Hessian for xi}
\eea
after the rescaling $\xi_\mu\to \sqrt{\tilde\a}\,\xi_\mu$.
In the $(\s,h,\varphi)$-basis, the Hessian for spin-0 scalar modes becomes
\bea
\frac{1}{2}
\pmat{\s & h& \varphi}\, \G^S \left(
\begin{array}{c}
\s \\
h \\
\varphi
\end{array}\right)
\equiv
\frac{1}{2}\pmat{\s & h& \varphi} \left(
\begin{array}{ccc}
\G_{\s\s} & \G_{\s h} & \G_{\s \phi} \\
\G_{h\s} & \G_{h h} & \G_{h \phi} \\
\G_{\phi\s} & \G_{\phi h} & \G_{\phi \phi}
\end{array}\right)
\left(
\begin{array}{c}
\s \\
h \\
\varphi
\end{array}\right),
\eea
where
\bea
&& \G^S = \small{
\left(
\begin{array}{ccc}
-\frac{3}{16} (\lich_0)^2 \left(\lich_0-\frac{\br}{3}\right) &
-\frac{3}{16} \lich_0 \left(\lich_0-\frac{\br}{3}\right) &
0 \\[2ex]
-\frac{3}{16} \lich_0 \left(\lich_0-\frac{\br}{3}\right) &
-\frac{3}{16} \left(\lich_0+\frac{\br}{3}\right) +\frac{1}{4}Z_N^{-1}U &
\frac{1}{2}Z_N^{-1/2} U' \sqrt{2\bar\rho} \\[2ex]
0 &
\frac{1}{2}Z_N^{-1/2} U' \sqrt{2\bar\rho} &
Z_\phi \lich_0+ M_H^2(\bar\rho)
\end{array}
\right)} \nn
&& \qquad\qquad \small{
+\frac{1}{\tilde\a} \left(
\begin{array}{ccc}
\frac{9}{16} \lich_0\left(\lich_0 -\frac{\br}{3}\right)^2 &
 \frac{3\b}{16}\lich_0\left(\lich_0-\frac{\br}{3}\right) &0\\[2ex]
\frac{3\b}{16}\lich_0\left(\lich_0-\frac{\br}{3}\right) & \frac{\b^2}{16}\lich_0 & 0 \\[2ex]
0 & 0 & 0
\end{array}
\right),}
\label{eq: Hessian for scalar fields}
\eea
where we have defined the effective Higgs scalar mass
\bea
M_H^2(\bar\rho)= U'(\bar\rho)+2U''(\bar\rho) \bar\rho.
\label{eq: field depend mass}
\eea

There are also determinants that have to be taken into account coming from changing field variables in the York decomposition:
\bea
\Det_{(1)}\left(\lich_1-\frac{\br}{2}\right)^{1/2}
\Det_{(0)}\left[\lich_0 \left(\lich_0-\frac{\br}{3}\right)\right]^{1/2}
\equiv J_\text{grav1} J_\text{grav0}.
\label{eq: Jaconians arising from decomposition}
\eea
See Appendix~\ref{york} for the derivation of this Jacobian.

Let us here consider properties of the fields appearing in our Hessian.
We first note that the scalar part in the gravity Hessian~\eqref{Hessg}
can be written as
\bea
-\frac{3}{32} s\left(\lich_0-\frac{\br}{3}\right)s -\frac{\br}{16} h^2,
\label{Hessgs}
\eea
where
\bea
s=\lich_0 \s+ h,
\label{eq: gauge-invariant variable}
\eea
is the gauge-invariant variable. On the other hand, if we use
the York decomposition, our gauge fixing function~\eqref{gff}
on the Einstein space~\eqref{es} becomes
\bea
f_\mu=-\left(\lich_1-\frac{\br}{2}\right)\xi_\mu-\nabla_\mu\left[\frac{3}{4}\left(\lich_0-\frac{\br}{3}\right)\s
+\frac{\b}{4} h\right].
\eea
We see that for the choice of $\b=3$, the scalar combination in this gauge fixing function becomes precisely
the gauge invariant variable $s$ modulo curvature term.
We will see that there is a singularity at $\b=3$ in various quantities in the following.
The origin of this singular behavior is that the longitudinal direction of the metric fluctuation is
not affected by the gauge fixing \cite{Gies:2015tca,Ohta:2016npm,Ohta:2016jvw}. In other words, the scalar modes in the gauge fixing function become
exactly the gauge-invariant combination and it does not fix the gauge for this value of $\b$. We also see related singularity in the longitudinal mode of the ghost field~\eqref{ghostkin}.

The gauge fixing function can be written as
\bea
f_\mu=-\left(\lich_1-\frac{\br}{2}\right)\xi_\mu-\frac{3-\b}{4}\nabla_\mu\left(\lich_0-\frac{\br}{3-\b} \right) \chi,
\eea
where
\bea\chi=\frac{(3\lich_0-\br)\s+\b h}{(3-\b)\lich_0-\br}
\eea
is a new degree of freedom. We find that $\chi$ transforms as $\s$. In the absence of matter, the last term in Eq.~\eqref{Hessgs} is
zero on shell, and the Hessian for gravity is written entirely in terms of the gauge-invariant variables $h^{TT}$ and $s$,
and the gauge fixing is entirely in terms of the gauge-variant fields $\xi$ and $\chi$.
If we consider $\b\to\infty$, we have $h=-\lich_0\chi$, and the gauge fixing strongly enforces the condition $\chi=0$, independently of $\a$.
This practically kills $h$, and setting $\a=0$ removes the gauge-variant field $\xi$, and then we have
only the contributions from gauge-invariant variables. Since this choice
sets $h=0$, this is called unimodular physical gauge~\cite{Percacci:2015wwa,Ohta:2015fcu}.
So the gauge choice $\a=0, \b=\infty$ may be an interesting physical gauge. Moreover, it is known that singularities in propagators of the metric fluctuation fields disappear in the pure gravity.
However, in the presence of scalar fields, we have seen that only the trace mode of the metric fluctuation makes contribution to the scalar potential, which vanishes in the $\b=\infty$ limit.
This implies that the metric fluctuations contribute only to $Z_\phi$ in this limit.

Another particular choice for the gauge fixing parameters may be $\tilde\alpha\to 0$ and $\beta=-1$~\cite{Wetterich:2016vxu,Wetterich:2016ewc,Pawlowski:2018ixd,Wetterich:2019zdo}.
For this choice, the gauge-variant modes $\xi_\mu$ and $\chi$ are removed by the gauge fixing and then the TT mode and the gauge invariant scalar modes \eqref{eq: gauge-invariant variable} remain as physical modes.

Note also that the choice $\tilde\alpha\to 0$ and $\beta=0$ is often employed. 
In this gauge fixing, $\beta=0$ keeps only $(\Gamma_\text{GF}^{(2)})_{\sigma\sigma}$ which gives the kinetic term of the $\sigma$ mode in the Landau gauge $\tilde\alpha\to 0$ [see Eq.~\eqref{eq: Hessian for scalar fields}] and thus the structure of the flow equations becomes simple.

\section{Flow equations}
\label{sec: Flow equations}

The flow equation for the system \eqref{eq: effective action in this work} can be schematically written as
\begin{align}
\pa_t \G_k &= \left. \frac{1}{2}\Tr \frac{\pa_t\mathcal R_k}{\G_k^{(2)}+ \mathcal R_k} \right|_{h^{TT}h^{TT}}
+\left. \frac{1}{2}\Tr \frac{\pa_t \mathcal R_k}{\G_k^{(2)}+ \mathcal R_k} \right|_{\xi\xi}
- \left. \Tr \frac{\pa_t\mathcal R_k}{\G_k^{(2)}+\mathcal R_k} \right|_{C^T C^T}
- \left. \frac{1}{2}\Tr \frac{\pa_t \mathcal R_k}{\G_k^{(2)}+\mathcal R_k} \right|_{J_\text{grav1}} \nn
&\qquad
+ \left. \frac{1}{2}\Tr \frac{\pa_t\mathcal R_k}{\G_k^{(2)}+\mathcal R_k} \right|_\text{scalar}
- \left. \Tr \frac{\pa_t \mathcal R_k}{\G_k^{(2)} + \mathcal R_k} \right|_{C^L C^L}
- \left. \frac{1}{2}\Tr \frac{\pa_t \mathcal R_k}{\G_k^{(2)}+\mathcal R_k} \right|_{J_\text{grav0}}.
\label{eq: all flow generators}
\end{align}
Here we employ the regulator such that Lichnerowicz Laplacians are replaced by $P_k=\lich+R_k(\lich)$, i.e.
\begin{align}
\mathcal R_k(\lich) = \Gamma_k^{(2)}(P_k) - \Gamma_k^{(2)}(\lich).
\end{align}
In this work, the flow generators in Eq.~\eqref{eq: all flow generators} are calculated using the optimized cutoff~\cite{Litim:2001up}
\bea
R_k(\lich)=(k^2-\lich) \theta(k^2-\lich).
\label{eq: optimized cutoff}
\eea
In the calculation of the contributions of
the spin-0 part, we should use the matrix form to take into account the mixing terms.

We define the dimensionless quantities by
\begin{align}
&\tilde U(\tilde\rho) = \frac{U(\bar\rho)}{k^4},&
&\tilde \rho = \frac{Z_\phi \bar\rho}{k^2},
\end{align}
as well as the dimensionless ratio
\bea
v(\tilde\rho) = \frac{2U(\bar\rho)}{M_P^2 k^2} = \frac{2 \tilde U(\tilde\rho)}{\tilde M_P^2},
\label{eq: v(rho)}
\eea
where the Planck mass and its dimensionless version are defined by
\begin{align}
&M_P^2 = 2Z_N,&
&\tilde M_P^2 = \frac{M_P^2}{k^2}.
\end{align}

We find the flow equation for the dimensionless effective scalar potential
\begin{align}
\partial_t \tilde U(\tilde\rho) &=-4\tilde U(\tilde\rho)+(2+\eta_\phi)\tilde\rho \partial_{\tilde\rho}\tilde U(\tilde\rho)
+ \frac{1}{16\pi^2}\ell_0^4(0)
+\frac{1}{16\pi^2}\ell_0^4(-\tilde M^2_s(\tilde \rho))\nn
&\qquad
+\frac{1}{8\pi^2}\left(1 -\tfrac{\eta_\phi}{6} \right)
\left[ 1 - 2\frac{4(3-\tilde\alpha)}{(3-\beta)^2\tilde M_P^2} \tilde U(\tilde\rho)\right]\ell_0^4(-\tilde M^2_s(\tilde \rho))\ell_0^4(\tilde M_H^2(\tilde\rho)),
\label{eq: flow equation of scalar potential}
\end{align}
where $\eta_\phi$ is the anomalous dimension of $\phi$:
\bea
\eta_\phi=-\frac{\pa_t Z_\phi}{Z_\phi}.
\eea
Details of the derivation are given in Appendix~\ref{app: sec: flow equation}.
Here $\tilde M_H^2(\tilde \rho)$ is the dimensionless version of the Higgs mass in Eq.~\eqref{eq: field depend mass} and we have also defined a $\tilde \rho$-dependent mass of scalar modes in the metric fluctuations:
\begin{align}
\tilde M^2_s(\tilde \rho)&=\frac{4(3-\tilde \alpha)}{(3-\beta)^2\tilde M_P^2}
\left[2\tilde U(\tilde\rho) - \frac{4 \tilde \rho (U'(\tilde \rho))^2}{1+\tilde M_H^2(\tilde \rho)} \right].
\label{eq: rho-dependent mass of scalar modes}
\end{align}
We have further introduced the shorthand threshold function
\begin{align}
\ell_p^{2n}(x)=\frac{1}{n!}\frac{1}{(1+x)^{p+1}}.
\label{eq: threshold function}
\end{align}
This function originates from loop integrals in the heat kernel expansion. See Appendix~\ref{app: sec: heat kernel} for the precise definition.

Next, we expand the potential in polynomials of $\tilde\rho$ around the origin $\tilde \rho=0$:
\bea
\tilde U(\tilde \rho)=\tilde V+\tilde m_H^2 \tilde \rho +\frac{\tilde \la}{2}\tilde\rho^2 +\cdots,
\label{eq: expansion of scalar potential}
\eea
where $\tilde m_H^2=M_H^2(\bar\rho=0)/k^2$.
Note that $\tilde V$ corresponds to the cosmological constant by $2 \tilde \Lambda=16\pi \tilde G_N \tilde V=\frac{2\tilde V}{\tilde M_P^2}$.
From this expansion~\eqref{eq: expansion of scalar potential}, we define the $\tilde\rho$-independent part of Eq.~\eqref{eq: v(rho)} and Eq.~\eqref{eq: rho-dependent mass of scalar modes},
respectively, by
\begin{align}
&v_0=\frac{2\tilde V}{\tilde M_P^2},&
&\tilde m_s^2=\frac{4(3-\tilde\alpha)}{(3-\beta)^2}v_0
=\left(\frac{4(3-\tilde\alpha)}{(3-\beta)^2\tilde M_P^2}\right) 2\tilde V,
\label{eq: mass of scalar modes}
\end{align}

\begin{figure}[t]
\begin{center}
\includegraphics[width=155mm]{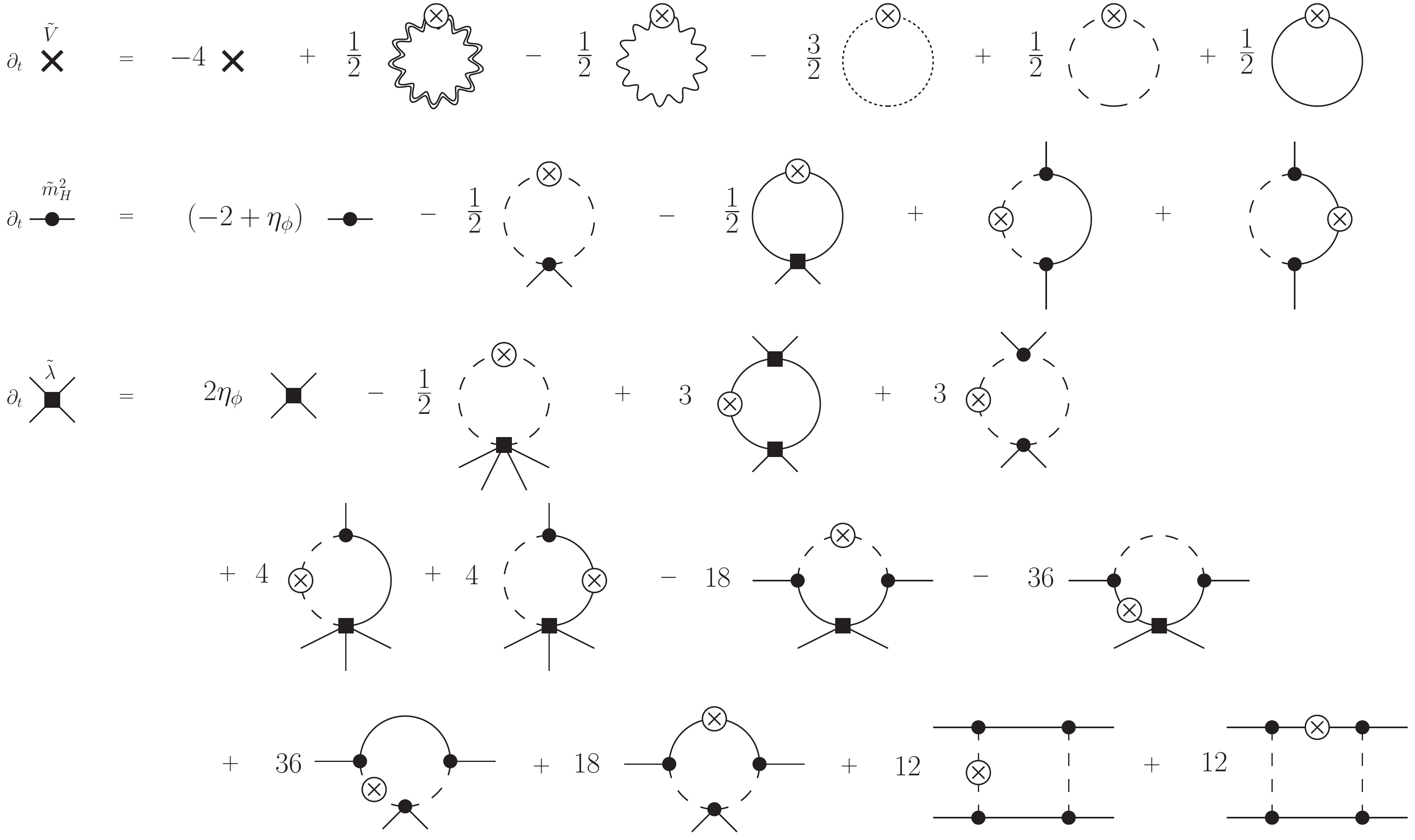}
\end{center}
\caption{
Diagrammatic representation for the beta functions of the cosmological constant, the scalar mass parameter and the quartic coupling denoted by a cross, a fat dot and a fat box, respectively.
The cross-circle stands for the cutoff insertion into the propagator, i.e. $\partial_t {\mathcal R}_k$. An external line is a background field $\bar\phi$. The double-wiggly, single-wiggly and dashed lines are the propagators for the spin-2 TT tensor,
spin-1 transverse vector and spin-0 scalar modes, respectively, while the dotted line is that of ghost fields and the Jacobians. The propagator of the scalar field fluctuation $\varphi$ is drawn as a solid line.
}
\label{fig:diagrams of beta functions}
\end{figure}

We find the beta functions for each coupling in the scalar potential \eqref{eq: expansion of scalar potential} as follows:
\begin{align}
\pa_t \tilde V &= -4\tilde V + \frac{1}{16\pi^2} \left[ \ell^4_0(0) + \ell^4_0(-\tilde m_s^2) + \left( 1 -\tfrac{\eta_\phi}{6}\right)\ell^4_0(\tilde m_H^2) \right],
\label{eq: beta function of cosmological constant}
\\
\pa_t \tilde m_H^2 &= (-2+ \eta_\phi + A )\tilde m_H^2 
-\left( 1 -\tfrac{\eta_\phi}{6}\right)\frac{3\tilde\lambda}{32\pi^2}\ell_1^2(\tilde m_H^2) \nn
&\qquad
-\frac{\tilde m_H^4}{8\pi^2}\frac{4(3-\tilde\alpha)}{(3-\beta)^2\tilde M_P^2} \left[  \ell_1^2(-\tilde m_s^2) \ell_0^2(\tilde m_H^2) + \left( 1-\tfrac{\eta_\phi}{6}\right) \ell_0^2(-\tilde m_s^2) \ell_1^2(\tilde m_H^2)   \right],
\label{eq: beta function of scalar mass parameter}
\\
\pa_t \tilde\la &= (2\eta_\phi + A)\tilde\lambda 
+\left( 1 -\tfrac{\eta_\phi}{6}\right)\frac{9\tilde\lambda^2}{16\pi^2}\ell_2^0(\tilde m_H^2)
+\frac{\tilde m_H^4}{4\pi^2}\left(\frac{4(3-\tilde\alpha)}{(3-\beta)^2\tilde M_P^2}\right)^2\ell_2^0(-\tilde m_s^2) \nn
&\qquad
- \frac{2\tilde m_H^2 \tilde\lambda}{4\pi^2}\frac{4(3-\tilde\alpha)}{(3-\beta)^2\tilde M_P^2}\left[ \ell_1^0(-\tilde m_s^2)\ell_0^0(\tilde m_H^2) + \left(1-\tfrac{\eta_\phi}{6} \right) \ell_0^0(-\tilde m_s^2)\ell_1^0(\tilde m_H^2)  \right]
\nn
&\qquad
-\frac{3\tilde m_H^4\tilde \lambda}{4\pi^2}\frac{4(3-\tilde\alpha)}{(3-\beta)^2\tilde M_P^2} \left[ \ell_1^0(-\tilde m_s^2)\ell_1^0(\tilde m_H^2)  + 2\left( 1-\tfrac{\eta_\phi}{6} \right)\ell_0^0(-\tilde m_s^2) \ell_2^0(-\tilde m_H^2) \right]\nn
&\qquad
-\frac{2 \tilde m_H^6}{4\pi^2}\left(\frac{4(3-\tilde\alpha)}{(3-\beta)^2\tilde M_P^2} \right)^2\left[ 2\ell^0_2(-\tilde m_s^2)\ell^0_0(\tilde m_H^2)  + \left( 1-\tfrac{\eta_\phi}{6} \right) \ell^0_1(-\tilde m_s^2)\ell^0_1(\tilde m_H^2) \right] \nn
&\qquad
+\frac{4\tilde m_H^8}{4\pi^2}\left(\frac{4(3-\tilde\alpha)}{(3-\beta)^2\tilde M_P^2} \right)^2 \left[ \ell^0_2(-\tilde m_s^2)\ell^0_1(\tilde m_H^2) + \left( 1-\tfrac{\eta_\phi}{6} \right)\ell^0_1(-\tilde m_s^2)\ell^0_2(\tilde m_H^2) \right] .
\label{eq: beta function of quartic coupling}
\end{align}
Here we have defined
\begin{align}
A =-\frac{\p}{\p U(\bar\rho)}(\p_t U(\bar\rho))\bigg|
_{\bar\rho=0}
=\frac{1}{16\pi^2}\frac{4(3-\tilde\alpha)}{(3-\beta)^2\tilde M_{P}^2}\ell_1^2(-\tilde m_s^2).
\label{metric induced anomalous dimension in exp. par.}
\end{align}
This quantity represents the anomalous dimension induced by the metric fluctuations contributing to the scalar potential, so hereafter we call it ``the metric-induced anomalous dimension".
In Appendix~\ref{app: sec: anomalous dimension of scalar field}, the anomalous dimension arising from the field renormalization of $\phi$ is found to be
\begin{align}
\eta_\phi 
&=\frac{5}{(4\pi)^2}\frac{1}{\tilde M_P^2}\ell_1^4(0)
+\frac{12\tilde\alpha}{(4\pi)^2}\frac{1}{\tilde M_P^2}\ell_2^6(0)
-\frac{1}{(4\pi)^2}\frac{4(3-\tilde\alpha)}{(3-\beta)^2\tilde M_P^2}\Bigg[
\frac{1}{4}\ell_1^2(-\tilde m_s^2) - \frac{9\tilde\alpha(3-\beta)^2}{(3-\tilde\alpha)^2}\ell_0^8(0) \nn
&\quad
+\frac{9(\tilde\alpha-\beta)^2}{(3-\tilde\alpha)^2}\left( \ell_1^8(-\tilde m_s^2) + 2 \ell_0^8(-\tilde m_s^2) \right)
\Bigg]
+\frac{6\tilde \alpha}{(4\pi)^2}\frac{1}{\tilde M_P^2}\left[
\ell_1^6(0)\ell_0^2(\tilde m_H^2)
 + \left(1 - \tfrac{\eta_\phi}{6} \right)\ell_1^6(\tilde m_H^2)\ell_0^2(0)
 \right]\nn
 &\quad
+\frac{2}{(4\pi)^2}\frac{4(3-\tilde\alpha)}{(3-\beta)^2\tilde M_P^2}\Bigg[ \ell_0^2(\tilde m_H^2)\Bigg(
\ell_1^6(-\tilde m_s^2)
+ \frac{18 ({\tilde \alpha} -\beta )}{(3-\tilde\alpha)} \left(  \ell_1^8(-\tilde m_s^2)+ \ell_0^8(-\tilde m_s^2) \right) \nn
&\quad
+ \frac{108({\tilde \alpha} -\beta )^2}{(3-{\tilde \alpha} )^2} \left( \ell_1^{10}(-\tilde m_s^2) +2 \ell_0^{10}(-\tilde m_s^2)  \right) 
- \frac{108{\tilde \alpha} (3-\beta)^2}{(3-{\tilde \alpha})^2}\ell_0^{10}(0)
 \Bigg) \nn
&\qquad
+\ell_1^2(\tilde m_H^2) \Bigg(
 \left( 1 -\tfrac{\eta_\phi}{8} \right)\ell_0^6(-\tilde m_s^2)
+  \left( 1-\tfrac{\eta_\phi}{10} \right) \frac{18 ({\tilde \alpha} -\beta )}{(3-\tilde\alpha )}\ell_0^8(-\tilde m_s^2) \nn
&\quad
+ \left( 1 -\tfrac{\eta_\phi}{12}  \right) \left(\frac{108({\tilde \alpha} -\beta )^2}{(3-{\tilde \alpha})^2}\ell_0^{10}(-\tilde m_s^2) - \frac{36 {\tilde \alpha} (3-\beta)^2 }{(3-{\tilde \alpha} )^2}\ell_0^{10}(0) \right)
\Bigg)
\Bigg] .
\label{anomalous dimension etaphi}
\end{align}
In Fig.~\ref{fig:diagrams of beta functions}, the beta functions \eqref{eq: beta function of cosmological constant}--\eqref{eq: beta function of quartic coupling} are diagrammatically displayed. In particular, the metric-induced anomalous dimension $A$ is generated by the one-loop diagram of scalar modes in the metric fluctuations which correspond to the fifth diagram on the rhs of $\p_t \tilde V$ in Fig.~\ref{fig:diagrams of beta functions}.
The second terms on the rhs of $\p_t\tilde m_H^2$ and $\p_t \tilde\lambda$ in Fig.~\ref{fig:diagrams of beta functions} also produce the effects with $A$. For higher dimensional operators, we have commonly 
\begin{align}
    A = \frac{\p}{\p \lambda_n}\left(-\frac{1}{2}~\vcenter{\hbox{\includegraphics[ width=15mm]{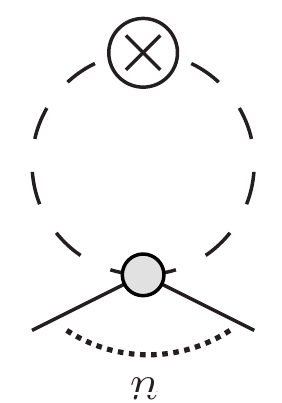}}}\right),
\end{align}
where $\lambda_n$ is a $\phi^n$-coupling which is denoted by a gray circle.
Diagrams contributing to $\eta_\phi$ are shown in Section~\ref{sec: Linear versus exponential parametrization} and Appendix~\ref{app: sec: anomalous dimension of scalar field}.

Note that if we can neglect $\eta_\phi$ and we redefine the Planck mass squared by
\begin{align}
\frac{4(3-\tilde\alpha )}{(3-\beta)^2\tilde M_P^2} \equiv \frac{1}{\widehat M_P^2},
\label{eq: repacement explict}
\end{align}
the explicit dependencies on the gauge fixing parameters in the scalar potential (including the cosmological constant) disappear. Therefore, using this redefined Planck mass, we can discuss the beta functions for the scalar potential without apparent gauge dependence.

On the other hand, the beta function of the (dimensionless) Planck mass squared reads
\begin{align}
\partial_t \tilde M_P^2 &= -2\tilde M_P^2 
+\frac{1}{8\pi^2}\frac{13}{3}\ell_0^2(0) +  \frac{1}{8\pi^2}\frac{2\beta}{3(3-\beta)}\ell_0^4(0)
- \frac{1}{48\pi^2}\left( 1- \tfrac{\eta_\phi}{4} \right)\ell_0^2(\tilde m_H^2)
-\frac{1}{48\pi^2}\ell_0^2(-\tilde m_s^2)
\nn
&\quad
+\frac{1}{48\pi^2} \frac{4\tilde\alpha-15}{3-\tilde\alpha} 
+\frac{1}{8\pi^2}\left[
\frac{1}{2(3-\tilde\alpha)} + \frac{3-\tilde\alpha}{(3-\beta)^2} -\frac{1}{3-\beta}
\right] \ell_1^2(-\tilde m_s^2).
\label{eq: flow equation of the Planck mass squared}
\end{align}
The singularity at $\tilde\alpha=3$ disappears if the definition of $\tilde m_s^2$ given in \eqref{eq: mass of scalar modes} is substituted. However, after the replacement \eqref{eq: repacement explict} is done, that singularity remains. We will discuss this issue in Section~\ref{sec: gauge dependence}.
Note that since the regulators for the metric fluctuations have no dependence on $M_P^2$ thanks to the rescaling $h_{\mu\nu}\to Z_N^{-1/2}h_{\mu\nu}$, there is no term with the anomalous dimension $\sim \p_t M_P^2/M_P^2$ on the rhs.

\section{Linear versus exponential parametrization}
\label{sec: Linear versus exponential parametrization}

In this section, we study the dependence of the metric-induced anomalous dimension $A$ on the fixed-point values of the Planck mass and the cosmological constant, and the pole structure of the propagator of the metric fluctuations, in the linear and exponential parametrizations.
We highlight the differences between the linear and exponential parametizations.

As pointed out in the introduction, the result using the standard linear split of the metric appears to have the problem of singularities in the cosmological constant due to poles in propagators of different modes involved in the metric fluctuations.
We would like to check if we can avoid the problem in the exponential parametrization and extract physical information by comparing the results with those using the linear split.
We will see that there is a pole even for the exponential parametrization in the presence of scalar fields.
Such a pole is considered to signal the presence of a phase transition, and should be studied beyond polynomial truncation~\cite{Berges:2000ew}.

However, as mentioned in the previous section, the redefinition of the Planck mass \eqref{eq: repacement explict} eliminates the apparent dependence of the gauge parameters from the beta function
of the effective potential in the absence of $\eta_\phi$ in the exponential parametrization.
Therefore, we can discuss the beta function of the Planck mass and the anomalous dimension of the scalar field without gauge dependence.
When we compare the results between the linear and exponential parametrizations, we take the gauge parameters $\tilde\a=0,~\b=1$ (de Donder gauge)~\cite{Narain:2009fy,Eichhorn:2017als}, $\tilde\a=0,~\b=-1$ (physical gauge)~\cite{Pawlowski:2018ixd,Wetterich:2019zdo} and $\tilde\a=0,~\b=\infty$ (unimodular gauge)~\cite{Percacci:2015wwa,Ohta:2015efa,Ohta:2015fcu}.
More precise analysis on the gauge dependence in the beta function of the Planck mass is presented in Section~\ref{sec: gauge dependence}.

Here we clarify a crucial difference in the contributions from the metric fluctuations to the dynamics of the scalar field between the linear and exponential parametrizations.
As will be seen later, the metric-induced anomalous dimensions $A$ and the scalar field $\eta_\phi$ play a central role in the deviation of the critical exponents of the couplings of the scalar field from their canonical scaling.
Let us consider the flow equation for the two-point function of the background field $\bar\phi$ in a flat spacetime background $\bar g_{\mu\nu}=\delta_{\mu\nu}$:
\begin{align}
\frac{\delta^2}{\delta \bar\phi(p)\delta \bar\phi(-p)}\partial_t\Gamma_k\bigg|_{\bar\phi=0}
= \int d^4x\,\left[-\partial_t Z_\phi p^2 + \p_t m_H^2 \right].
\end{align}
To make the discussion simple, we set $\lambda=0$.

In the exponential parametrization, the flow equations for $Z_\phi$ and $m_H^2$ are represented diagrammatically as 
\begin{align}
    \p_t Z_\phi &\sim \frac{d}{dp^2}\left(
-\tfrac{1}{2}\vcenter{\hbox{\includegraphics[bb=0 0 145 97, width=22mm]{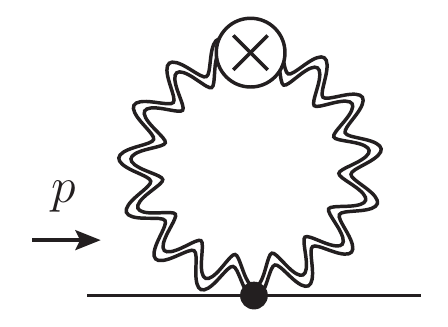}}}
-\tfrac{1}{2}\vcenter{\hbox{\includegraphics[bb=0 0 145 97, width=22mm]{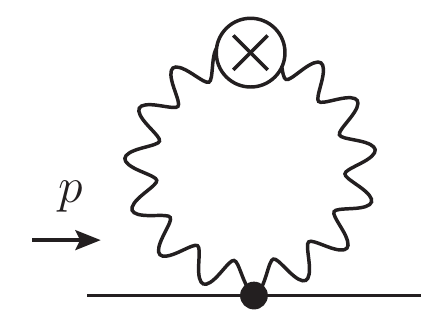}}}
-\tfrac{1}{2}\vcenter{\hbox{\includegraphics[bb=0 0 145 91, width=23mm]{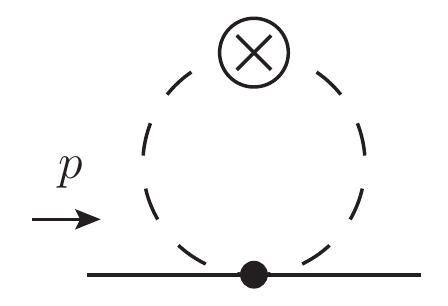}}}
\right.\nn
&\quad
\left.\left.
+\vcenter{\hbox{\includegraphics[bb=0 0 137 71, width=25mm]{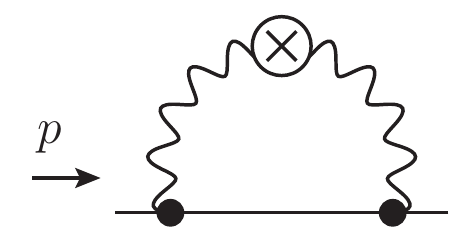}}}
+\vcenter{\hbox{\includegraphics[bb=0 0 137 72, width=25mm]{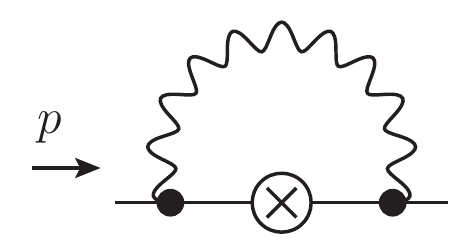}}}
+\vcenter{\hbox{\includegraphics[bb=0 0 137 71, width=25mm]{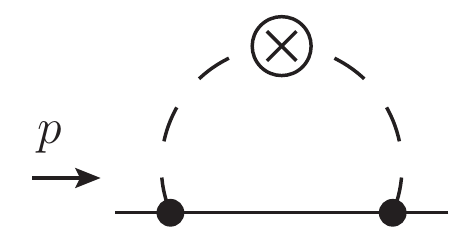}}}
+\vcenter{\hbox{\includegraphics[bb=0 0 137 67, width=25mm]{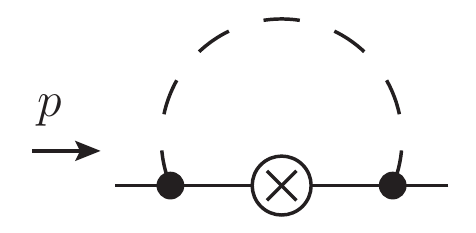}}}
\right)\right|_{p^2=0},
\label{eq: diagrammatic flow equation of Zphi in exponential}
\\[2ex]
\p_t m_H^2 &\sim 
\left.
-\tfrac{1}{2}\vcenter{\hbox{\includegraphics[bb=0 0 145 91, width=22mm]{tadpole_spin0}}}
+\vcenter{\hbox{\includegraphics[bb=0 0 137 71, width=22mm]{sunset_spin02}}}
+\vcenter{\hbox{\includegraphics[bb=0 0 137 67, width=23mm]{sunset_spin01}}}
\right|_{p^2=0},
\label{eq: diagrammatic flow equation of mass in exponential}
\end{align}
where the double-wiggly, single-wiggly and dashed lines denote the TT, spin-1 transverse vector and spin-0 modes in the metric fluctuations, respectively, while the solid line represents the scalar field.
The fat-dot and cross-circle stand for $Z_\phi$ or $m_H^2$ and $\p_t {\mathcal R}_k$, respectively. In particular, the first diagram on the rhs of Eq.~\eqref{eq: diagrammatic flow equation of mass in exponential} generates the metric-induced anomalous dimension $A$.
Note that the sunset diagrams (like the second line of Eq.~\eqref{eq: diagrammatic flow equation of Zphi in exponential}) with the TT mode vanish due to its transverse property.
We stress here that all modes in the metric fluctuations contribute to $Z_\phi$, whereas the scalar potential receives corrections from only scalar modes.
As a consequence the apparent gauge dependence may be eliminated by the redefinition of the Planck mass \eqref{eq: repacement explict}.

On the other hand, the use of the linear parametrization changes to the opposite situation:
\begin{align}
    \p_t Z_\phi &\sim \frac{d}{dp^2}\left.\left(
\vcenter{\hbox{\includegraphics[bb=0 0 137 71, width=23mm]{sunset_spin12}}}
+\vcenter{\hbox{\includegraphics[bb=0 0 137 72, width=23mm]{sunset_spin11}}}
+\vcenter{\hbox{\includegraphics[bb=0 0 137 71, width=23mm]{sunset_spin02}}}
+\vcenter{\hbox{\includegraphics[bb=0 0 137 67, width=23mm]{sunset_spin01}}}
\right)\right|_{p^2=0},
\label{eq: diagrammatic flow equation of Zphi in linear}
\\
\p_t m_H^2 &\sim 
-\tfrac{1}{2}\vcenter{\hbox{\includegraphics[bb=0 0 145 97, width=22mm]{tadpole_tt}}}
-\tfrac{1}{2}\vcenter{\hbox{\includegraphics[bb=0 0 145 97, width=22mm]{tadpole_spin1}}}
-\tfrac{1}{2}\vcenter{\hbox{\includegraphics[bb=0 0 145 91, width=22mm]{tadpole_spin0}}}
+\vcenter{\hbox{\includegraphics[bb=0 0 137 71, width=23mm]{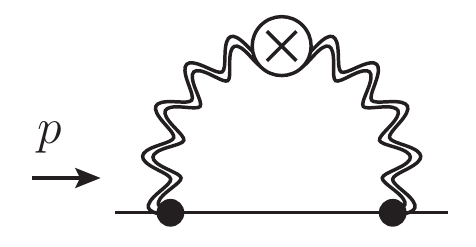}}}
+\vcenter{\hbox{\includegraphics[bb=0 0 137 72, width=23mm]{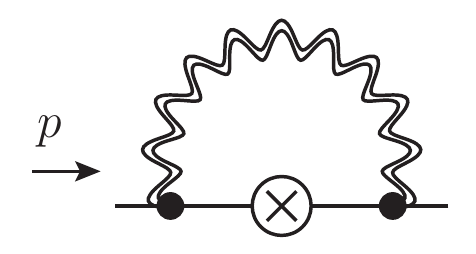}}}
\nn[2ex]
&\qquad
\left.
+\vcenter{\hbox{\includegraphics[bb=0 0 137 71, width=23mm]{sunset_spin12}}}
+\vcenter{\hbox{\includegraphics[bb=0 0 137 72, width=23mm]{sunset_spin11}}}
+\vcenter{\hbox{\includegraphics[bb=0 0 137 71, width=23mm]{sunset_spin02}}}
+\vcenter{\hbox{\includegraphics[bb=0 0 137 67, width=23mm]{sunset_spin01}}}
\right|_{p^2=0}.
\label{eq: diagrammatic flow equation of mass in linear}
\end{align}
The first three tadpole diagrams on the rhs of Eq.~\eqref{eq: diagrammatic flow equation of mass in linear} yield finite corrections to the metric-induced anomalous dimension $A$.
An important fact is that the tadpole diagrams do not contribute to the flow equation of $Z_\phi$ in the linear parametrization.\footnote{The diagrams with the spin-1 transverse mode is proportional to $\tilde\alpha$, so the Landau gauge $\tilde\alpha\to0$ erases their contributions and then $\eta_\phi$ could be written as an apparent gauge invariant form, i.e. no $\beta$ dependence, within contributions from scalar modes. } 

We thus expect that the exponential parametrization produces a larger value of $\eta_\phi$ and a smaller value of $A$ than those in the linear parametrization case.
Even though the exponential parametrization appears to be appropriate because the gauge dependence can be absorbed into the redefinition of the Planck mass, a larger value of $\eta_\phi$ may cause a potential problem in the present local potential approximation, which requires $\eta_\phi=0$, and the derivative expansion for the effective action may not be valid.
So let us first find the fixed points and critical exponents in our setup and compare the results with the linear parametrization~\cite{Eichhorn:2017als}, and then check how the anomalous dimensions $A$ and $\eta_\phi$ affect the results. Eventually we would like to see if there is any optimal choice of $\b$ so as to make $\eta_\phi$ small.

\subsection{Fixed point}

We first study how the fixed points for the Planck mass and the cosmological constant change in the exponential parametrization~\eqref{exp}.
To this end, we suppose hereafter that the Gaussian matter fixed point, namely matter couplings have only the trivial fixed point, $\tilde m_{H*}^2=\tilde\lambda_*=0$.

The fixed point of $\tilde M_P^2$ and $\tilde V$ in the linear parametrization with $\tilde\a=0,~\b=1$ is found to be~\cite{Eichhorn:2017als}
\begin{align}
&\tilde M_{P*}^2 = 0.03366,&
&\tilde G_{N*} = 1.182,&
&\tilde V_* =0.005420,
\end{align}
whereas in our exponential parametrization, we find
\begin{align}
&\tilde M_{P*}^2 =0.03567,&
&\tilde G_{N*} = 1.1154,&
&\tilde V_* =0.003539,&
&(\text{for $\tilde\a=0,~\b=1$}),
\label{eq: fixed point in original Planck mass: beta=1}
\end{align}
\begin{align}
&\tilde M_{P*}^2 =0.020283,&
&\tilde G_{N*} = 1.9617,&
&\tilde V_* =0.0025595,&
(\text{for $\tilde\a=0,~\b=-1$}),
\label{eq: fixed point in original Planck mass: beta=-1}
\end{align}
\begin{align}
&\tilde M_{P*}^2 =0.018998,&
&\tilde G_{N*} = 2.0944,&
&\tilde V_* =0.0023747,&
(\text{for $\tilde\a=0,~\b=\infty$}).
\label{eq: fixed point in original Planck mass: beta=infty}
\end{align}
In terms of the redefined Planck mass \eqref{eq: repacement explict},
one has, for $\tilde\a=0,~\b=1$
\begin{align}
&\widehat M_{P*}^2 = 0.01189,&
&\widehat G_{N*} = 3.3464,&
&\tilde V_* =0.003539,
\label{eq: redefined case fixed point; alpha:0 beta:1}
\end{align}
and for $\tilde\a=0,~\b=-1$,
\begin{align}
&\widehat M_{P*}^2 = 0.027044,&
&\widehat G_{N*} = 1.4713,&
&\tilde V_* =0.0025595.
\label{eq: redefined case fixed point; alpha:0 beta:-1}
\end{align}
We do not find any reliable fixed point for $\tilde\a=0,~\b=\infty$, because the beta function is proportional to $\beta$. 
We will discuss this problem in detail in Section~\ref{sec: gauge dependence}.
The fixed points \eqref{eq: redefined case fixed point; alpha:0 beta:1} and \eqref{eq: redefined case fixed point; alpha:0 beta:-1} are related
to \eqref{eq: fixed point in original Planck mass: beta=1} and \eqref{eq: fixed point in original Planck mass: beta=-1}, respectively, via the relation \eqref{eq: repacement explict}.

We next assume that a fixed point of the Planck mass exists and treat its fixed-point value as a constant parameter. Note that a small value of the Planck constant corresponds to a strong interaction
of gravity.

\begin{figure}
\begin{center}
\includegraphics[width=77mm]{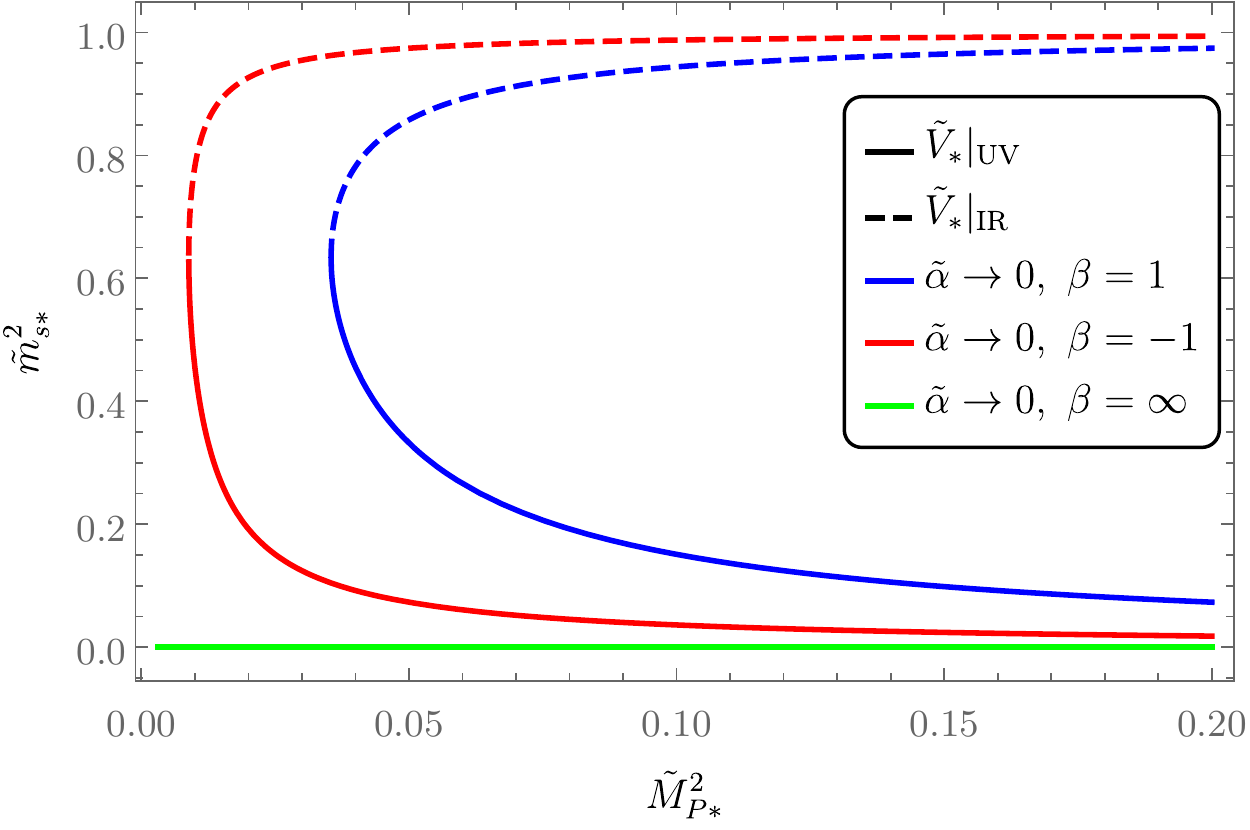}
\hs{2}
\includegraphics[width=77mm]{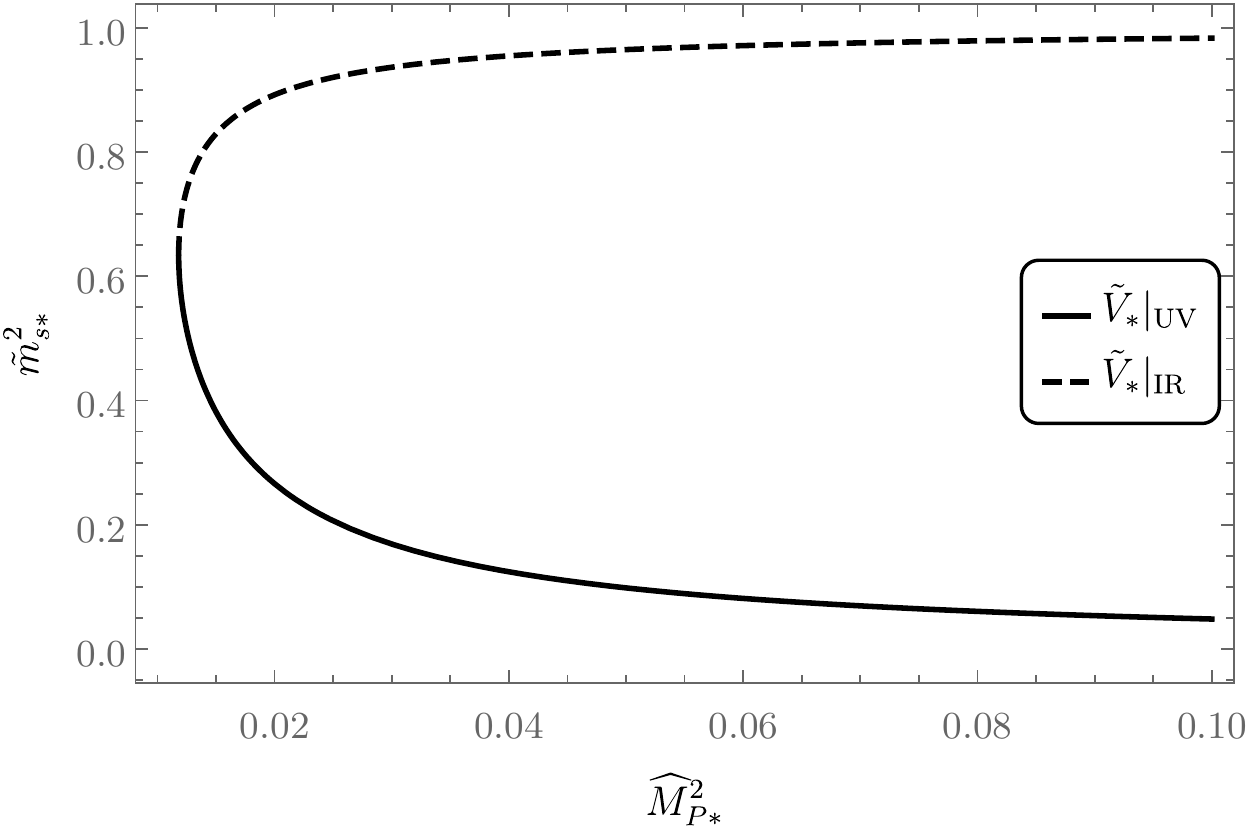}
\end{center}
\caption{The fixed-point value of the mass $\tilde m_{s*}^2$ as a function of the Planck mass squared.
Left: $\tilde m_{s*}^2$ as a function of $\tilde M_{P*}^2$ with different gauge parameter choices. Right: $\tilde m_{s*}^2$ as a function of $\widehat M_{P*}^2$.
The solid and dashed lines show the $\tilde m_{s*}^2$ with $\tilde V_{*}|_\text{UV}$ and $\tilde V_{*}|_\text{IR}$ defined in Eq.~\eqref{eq: fixed point of Lambda}, respectively.
}
\label{fig:cosmp1}
\end{figure}

With the redefinition \eqref{eq: repacement explict}, no dependence on the gauge parameter appears in the beta function of the cosmological constant since it does not have $\eta_\phi$.
For vanishing matter fixed points ($\tilde m_{H*}^2=\tilde\lambda_*=0$), we find from Eq.~\eqref{eq: beta function of cosmological constant} fixed points of $\tilde V$:
\begin{align}
\tilde V_{*}|_\text{UV}&= \frac{1+32\pi^2\widehat M_{P*}^2- \sqrt{1-128\pi^2\widehat M_{P*}^2 +1024\pi^4 \widehat M_{P*}^4}}{128\pi^2},\nn[1ex]
\tilde V_{*}|_\text{IR}&= \frac{1+32\pi^2\widehat M_{P*}^2+ \sqrt{1-128\pi^2\widehat M_{P*}^2 +1024\pi^4 \widehat M_{P*}^4}}{128\pi^2}.
\label{eq: fixed point of Lambda}
\end{align}
In Fig~\ref{fig:cosmp1}, we plot $\tilde m_{s*}^2=2\tilde V_*(4(3-\tilde\alpha)/(3-\beta)^2\tilde M_P^2)=2\tilde V_{*}/\widehat M_{P*}^2$ instead of $\tilde V_{*}$.
There also exist fixed points for smaller values of $\tilde M_{P*}^2$ or $\widehat M_{P*}^2(\leq\frac{2-\sqrt{3}}{32\pi^2}\simeq 0.00085)$; however, the fixed point values of $\tilde m^2_{s}$ there exceed 1 and thus we should exclude them.
One can have these fixed points of $\tilde V$ written in terms of the original Planck mass $\tilde M_P^2$ by using Eq.~\eqref{eq: repacement explict}.

We see from Fig~\ref{fig:cosmp1} that for larger values of $\widehat M_{P*}^2$, the mass parameter $\tilde m_{s*}^2|_\text{IR}$ converges to $1$ which is a pole of the threshold functions $\ell_p^{2n}(-\tilde m_s^2)$ and then $\tilde V_*|_\text{IR}$ corresponds to the IR fixed point, while $\tilde V_{*}|_\text{UV}$ is the UV fixed point.
In IR regimes, the flow of $\tilde m_s^2$ may come close to $\tilde m_{*}^2|_\text{IR}$ which would induce large effects of the metric fluctuations on interactions including the cosmological constant itself.
Since the metric fluctuations tend to make the cosmological constant smaller, it is argued in Refs.~\cite{Wetterich:2017ixo,Wetterich:2018qsl} that such an IR instability around a pole in the propagator could play a key role in realizing the tiny value of the cosmological constant.

\subsection{Critical exponents}
\label{sec: Critical exponent of the scalar effective potential}

We evaluate the critical exponents for couplings in the scalar potential.
Suppose here that the cosmological constant and the Planck mass have a non-trivial fixed point, while only the Gaussian fixed point is found for the scalar mass parameter and the quartic coupling.

The critical exponents are defined by
\begin{align}
    \theta_i = - \text{eig}\left(\frac{\partial\beta_i}{\partial g_j}\right)\bigg|_{\tilde g=\tilde g_*},
\end{align}
where $\tilde g_i=\{\tilde M_P^2 ,\tilde V, \tilde m_H^2, \tilde\lambda \}$ and ``eig" denotes evaluating eigenvalues of a matrix.

\begin{figure}
\begin{center}
\includegraphics[width=77mm]{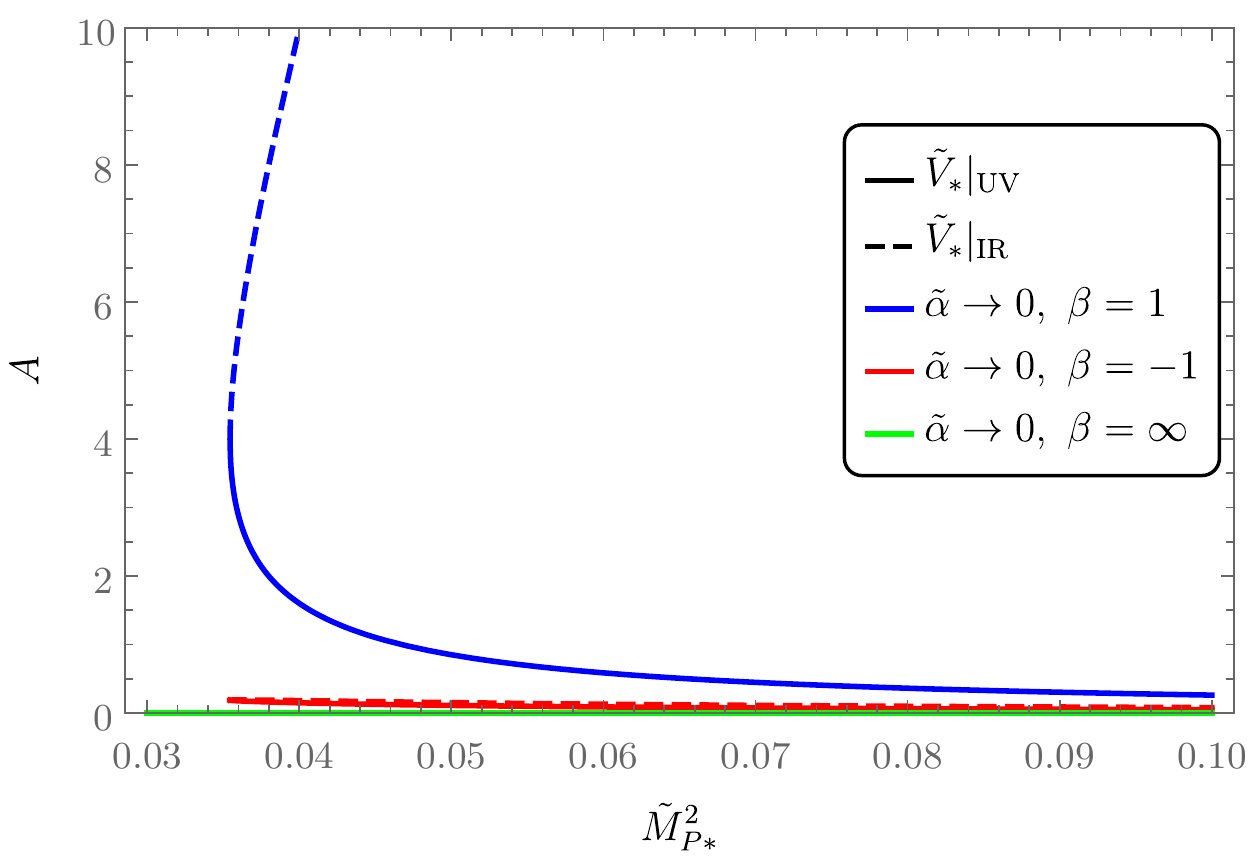}
\hs{2}
\includegraphics[width=77mm]{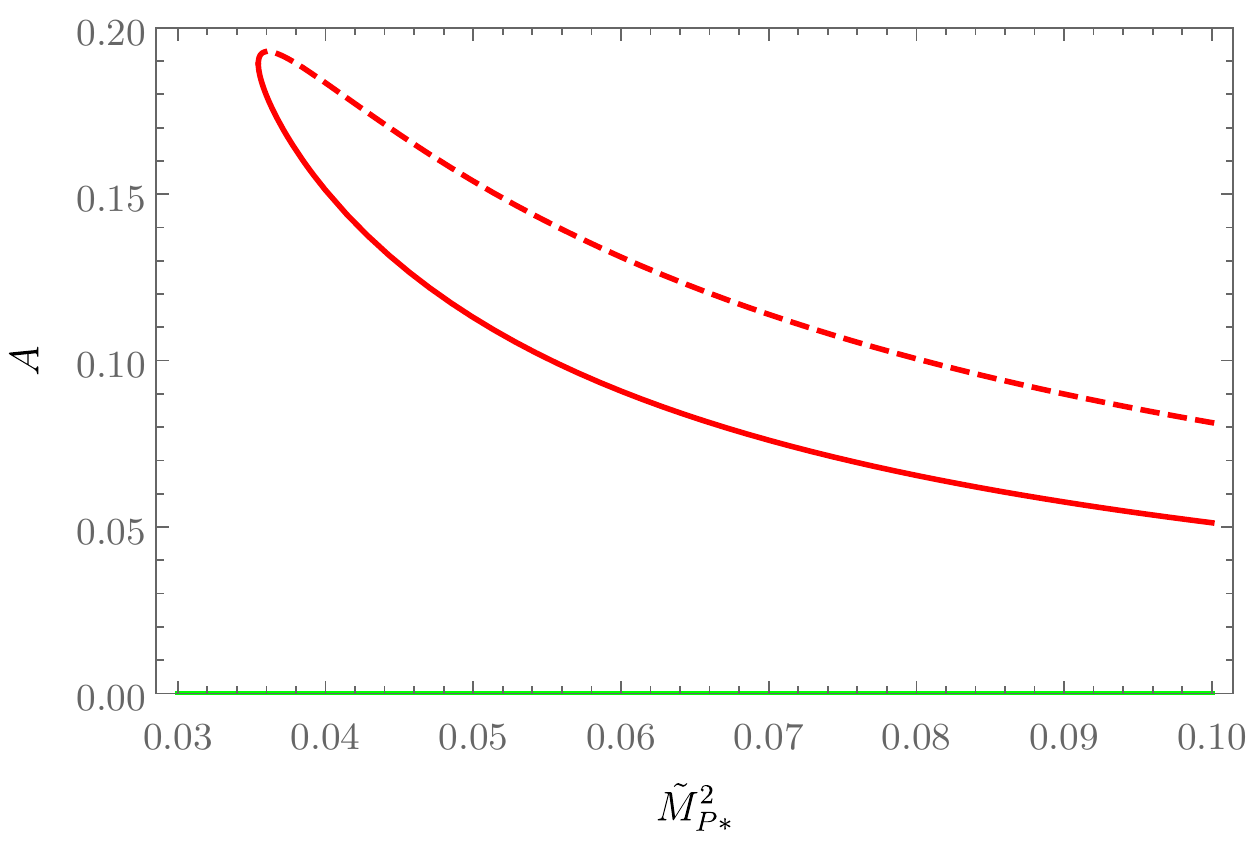}
\end{center}
\caption{The metric-induced anomalous dimension $A$ as a function
of $\tilde M_{P_*}$. The solid line corresponds to $V_*|_{\rm UV}$.
For the fixed point of $\tilde V$, we use Eq.~\eqref{eq: fixed point of Lambda}. The solid line corresponds to $V_*|_\text{UV}$ and dashed line to $V_*|_\text{IR}$. The rhs panel zooms in on the region of smaller $A$.}
\label{fig:critcosmp1_original}
\end{figure}

\begin{figure}
\begin{center}
\includegraphics[width=120mm]{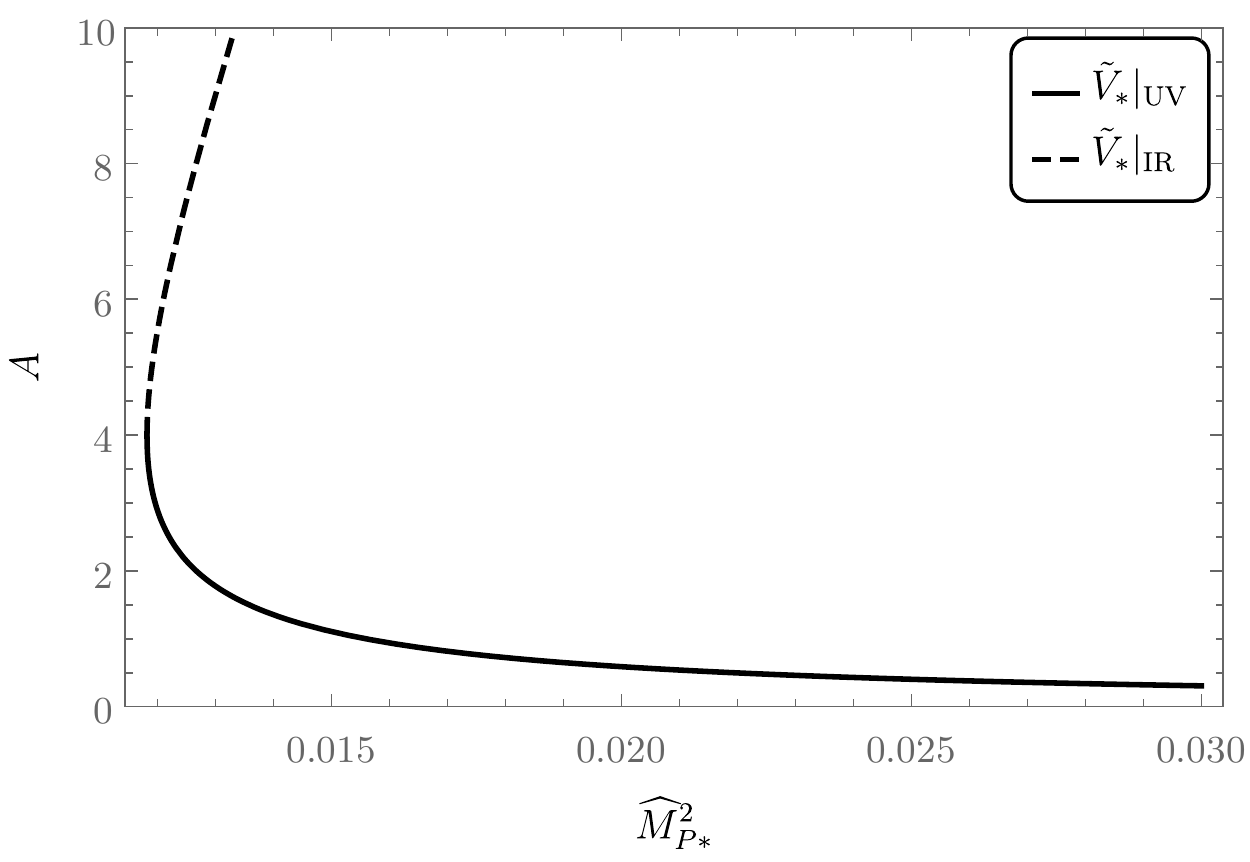}
\end{center}
\caption{The metric-induced anomalous dimension $A$ as a function
of $\widehat M_{P_*}$. For the fixed point of $\tilde V$, we use Eq.~\eqref{eq: fixed point of Lambda}.}
\label{fig:critcosmp1}
\end{figure}

Taking only the diagonal parts in the stability matrix $\frac{\partial\beta_i}{\partial g_j}$ into account, the critical exponents of the cosmological constant, the scalar mass parameter and the quartic coupling is approximately given, respectively, by
\begin{align}
&\theta_{\tilde V} \simeq -\frac{\partial \beta_{\tilde V}}{\partial \tilde V}\bigg|_{\tilde g=\tilde g_*}
= 4- A,
\label{eq: approximated critical exponent of CC}
\\[2ex]
&\theta_{\tilde m_H^2} \simeq -\frac{\partial \beta_{\tilde m_H^2}}{\partial \tilde m_H^2}\bigg|_{\tilde g=\tilde g_*} = 2 -\eta_\phi - A,
\label{eq: approximated critical exponent of scalar mass}
\\[2ex]
&\theta_{\tilde\lambda} \simeq -\frac{\partial \beta_\lambda}{\partial \tilde \lambda}\bigg|_{\tilde g=\tilde g_*}=-2\eta_\phi -A.
\label{eq: approximated critical exponent of quartic}
\end{align}
Note here that for the critical exponent of the beta function, one computes
\begin{align}
\frac{\partial}{\partial \tilde V}\frac{1}{16\pi^2}\ell_0^4(-\tilde m_s^2)\bigg|_{\tilde g=\tilde g_*} =A.
\end{align}
In the case of the exponential parametrization, the TT spin-2 tensor does not couple to the scalar potential, so it does not induce the anomalous dimension $A$ as can be seen in Eq.~\eqref{metric induced anomalous dimension in exp. par.}.
The critical exponent of the cosmological constant evaluated by Eq.~\eqref{eq: approximated critical exponent of CC}, however, may be insufficient because the mixing effects with the Planck mass cannot be neglected. More specifically, one should calculate 
\begin{align}
\theta_{\tilde M_P^2,\,\tilde V} = -\text{eig}\pmat{
\frac{\p \beta_{\tilde M_P^2}}{\p \tilde M_P^2} && \frac{\p \beta_{\tilde M_P^2}}{\p \tilde V}\\[2ex]
\frac{\p \beta_{\tilde V}}{\p \tilde M_P^2} && \frac{\p \beta_{\tilde V}}{\p\tilde V} 
}\Bigg|_{\substack{\tilde M_P^2=\tilde M_{P*}^2\\ \tilde V= \tilde V_{*}}}.
\label{eq: stabilit matrix of PM and CC}
\end{align}
We plot the metric-induced anomalous dimension \eqref{metric induced anomalous dimension in exp. par.} as a function of $\tilde M_{P*}^2$ in Fig.~\ref{fig:critcosmp1_original} and as a function of $\widehat M_{P*}^2$ in Fig.~\ref{fig:critcosmp1}.

When $\eta_\phi$ can be neglected, the critical exponents of couplings in the scalar potential are given as a constant shift $-A$ from their canonical dimensions.
In particular, the critical exponent of the scalar mass parameter becomes zero (or equivalently $A=2$) at the fixed point value of the Planck mass,
\begin{align}
\widehat M_{P*}^2 =\frac{1}{8\pi^2} \simeq 0.0126651.
\end{align}
The critical exponent of the cosmological constant vanishes when $A=4$ for which
\begin{align}
\widehat M_{P*}^2 =\frac{2+\sqrt{3}}{32 \pi ^2} \simeq 0.0118167.
\end{align}
At this value, one has $\tilde V_{*}|_\text{UV}=\tilde V_{*}|_\text{IR}$.

Note here that using the linear parametrization produces the metric-induced anomalous dimension $A$ given by
\begin{align}
A\big|_\text{linear}&= \frac{1}{24\pi^2 \tilde M_{P*}^2}\left[ 5\ell_1^2(-v_{0*})
+3\tilde\alpha\ell_1^2(-\tilde\alpha v_{0*})
+ \frac{2\left(3-\beta^2\right) }{(3-\beta)^2} \ell_1^2(-\tilde m_{L*}^2) \right.\nn
&\quad
\left. 
+\left( 4\tilde\alpha \frac{(3-\beta )^2 + 2 \left(3-\beta^2\right) v_{0*}^2 -4(3-\beta )^2v_{0*}}{(3-\beta)^4} 
+\frac{16\alpha^2 v_{0*}^2}{(3-\beta)^4} 
\right)
\ell_1^2(-\tilde m_{L*}^2)
\right],
\label{metric induced anomalous dimension in linear par.}
\end{align}
where we have defined the mass of scalar modes in metric fluctuations:
\begin{align}
\tilde m_L^2= \frac{2(3-\beta ^2)}{(3-\beta)^2}v_{0} +\frac{4\tilde\alpha }{(3-\beta)^2}(1-v_0)v_{0}.
\end{align}
The first and second terms in Eq.~\eqref{metric induced anomalous dimension in linear par.} are contributions from  the TT spin-2 tensor (i.e. graviton) and the transverse spin-1 vector, respectively, and the scalar modes of the metric fluctuations yield the remaining terms.

We now give numerical results for the critical exponents in the case of $\eta_\phi=0$.
At the fixed points of the original Planck mass squared $\tilde M_P^2$, the critical exponents are obtained for $\tilde\alpha\to0$, $\beta=1$ with Eq.~\eqref{eq: fixed point in original Planck mass: beta=1} as
\begin{align}
&\theta_{1,2}= 2.59539 \pm 2.12893 i,&
&\theta_3= -1.25264,&
&\theta_4= -3.25264,
\label{eq: critical exponents in orgial Mp in alpha0 beta1}
\end{align}
for $\tilde\alpha\to0$, $\beta\to-1$ with Eq.~\eqref{eq: fixed point in original Planck mass: beta=-1} as
\begin{align}
&\theta_{1}= 3.6402,&
&\theta_{2}= 2.020,&
&\theta_3= 1.6437,&
&\theta_4= -0.35628,
\label{eq: critical exponent for beta=-1}
\end{align}
and for $\tilde\alpha\to0$, $\beta\to\infty$ with Eq.~\eqref{eq: fixed point in original Planck mass: beta=infty} as
\begin{align}
&\theta_{1}= 4,&
&\theta_{2}= 2,&
&\theta_3= 2,&
&\theta_4= 0,
\label{eq: critical exponent for beta=infty}
\end{align}
where $\theta_1$ and $\theta_2$ correspond to the critical exponents of the cosmological constant and the Planck mass squared, respectively, whereas $\theta_3$ and $\theta_4$ are those of the scalar mass parameter and the quartic coupling, respectively.

When the redefined Planck mass is used, we have found a fixed point for $\tilde\alpha\to 0$, $\beta=1$ and $\tilde\alpha\to 0$, $\beta=-1$ as in Eqs.~\eqref{eq: redefined case fixed point; alpha:0 beta:1} and \eqref{eq: redefined case fixed point; alpha:0 beta:-1} at which the same critical exponents as Eqs.~\eqref{eq: critical exponents in orgial Mp in alpha0 beta1} and \eqref{eq: critical exponent for beta=-1} are observed.

To understand these results, several comments are in order:

First, the negative critical exponents in Eq.~\eqref{eq: critical exponents in orgial Mp in alpha0 beta1} result from a large value of $A$ and all scalar interactions including the scalar mass parameter become irrelevant.
This is because the fixed point value \eqref{eq: fixed point in original Planck mass: beta=1} and $\tilde\alpha\to 0$, $\beta=1$ yields $\tilde m_s^2\simeq 0.59533$ which locates close to the pole of the threshold function. Indeed, for this value, one has $\ell_1^2(-\tilde m_s^2)\simeq 6.1071$.
This casts a doubt to the validity of the result.

Second, the critical exponents for couplings of the scalar field for $\tilde\alpha\to0$, $\beta\to\infty$ remain the same as the canonical ones~\eqref{eq: critical exponent for beta=infty}. This comes from the fact that the $\beta$-dependence appears in denominators of terms in the beta function of the scalar potential \eqref{eq: flow equation of scalar potential} in terms of the original Planck mass squared, so the limit $\beta\to \infty$ (and $\tilde\alpha\neq 3$) suppresses the metric fluctuations.
This results in no metric-induced anomalous dimension even when the Planck mass has a finite fixed point value.

Third, if we use the redefined Planck mass squared, the beta function of the scalar potential has no explicit gauge-parameter dependence in contract to the case of the original Planck mass squared and the redefinition gives a finite $A$ except for $\widehat M_P^2\to\infty$. In this case we get the same result as \eqref{eq: critical exponent for beta=-1}.

Finally, the discrepancy in $\beta\to \infty$ discussed above could be understood by taking $\eta_\phi$ into account.
On the one hand, the metric fluctuation decouples from the scalar potential. On the other hand, the kinetic term of the scalar field gets large contribution from the metric fluctuation,
resulting in the anomalous dimension with the redefined Planck mass proportional to polynomials of $\beta$: $\eta_\phi \propto \frac{1}{\tilde M_P^2}=\frac{(3-\beta)^2}{4(3-\tilde\alpha)\widehat M_P^2}$.
Thus, we expect that in the limit $\beta\to \infty$, the anomalous dimension $\eta_\phi$ diverges. Indeed, such a divergent behavior is observed in the linear parametrization case as well.
This means that it is better that $|\b|$ is not too large.

\subsection{Effects of the anomalous dimension}

Let us now discuss the effects of the anomalous dimension $\eta_\phi$ of the scalar field. In order for the derivative expansion to be valid for the effective action, $|\eta_\phi|<1$ should be satisfied.
We might allow somewhat larger $|\eta_\phi|$ within the current approximation because the loop effect of the scalar field on the Planck mass squared and the scalar potential has suppression factors
such as $\eta_\phi/4$ and $\eta_\phi/6$; see Eqs.~\eqref{eq: flow equation of scalar potential} and \eqref{eq: flow equation of the Planck mass squared}.
In the linear parametrization, $\eta_\phi$ is of order  $0.1$~\cite{Eichhorn:2017als}. In particular, the choice $\beta=1$ in the Landau gauge $\tilde\alpha \to 0$ gives the vanishing $\eta_\phi$ in the case of the linear parametrization.
In the exponential parametrization, however, the TT mode, which is independent of the gauge parameters, may make a significant contribution to $\eta_\phi$.

\begin{figure}[t]
\begin{center}
\includegraphics[width=120mm]{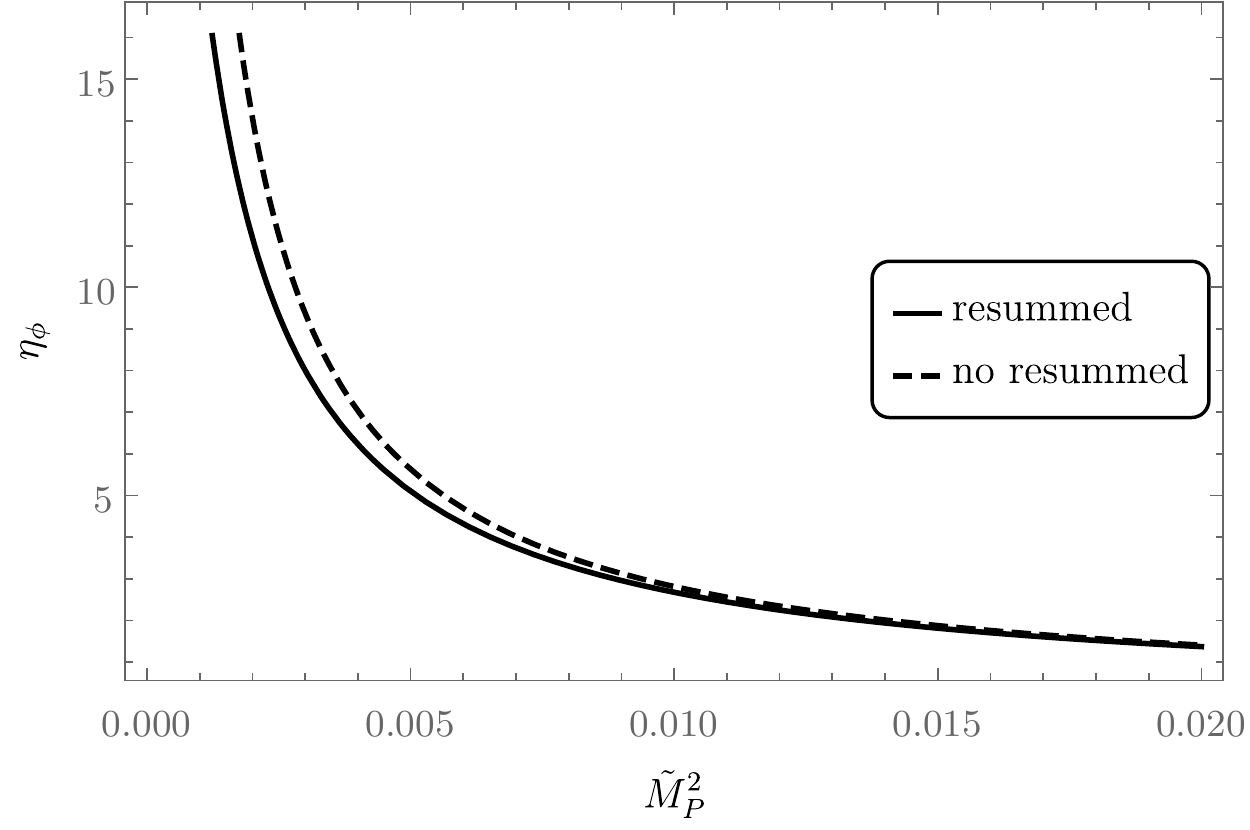}
\end{center}
\caption{The anomalous dimension of the scalar field $\eta_\phi$ as a function of $\tilde M_P^2$. We use $\tilde\alpha\to0$, $\beta=-1$ and $\tilde V=0$ for the presentation.}
\label{fig:anomalous_eta_phi}
\end{figure}

The rhs of Eq.~\eqref{anomalous dimension etaphi} has the $\eta_\phi$ dependence arising from the scalar field propagator.
Here, we consider two cases: (i) setting $\eta_\phi=0$ in the rhs of Eq.~\eqref{anomalous dimension etaphi}; (ii) solving for $\eta_\phi$ in Eq.~\eqref{anomalous dimension etaphi}. We call the former (latter) ``no-resummed" (``resummed") $\eta_\phi$.
The anomalous dimension $\eta_\phi$ as a function of $\tilde M_P^2$ is shown in Fig.~\ref{fig:anomalous_eta_phi} where we take $\tilde\alpha\to 0$, $\beta=-1$ and set $\tilde V=0$ as an example.
The resummed $\eta_\phi$ becomes relatively smaller than no-resummed one. The larger the value of $\eta_\phi$ is, the larger is the difference between them. For a reasonable value of $\eta_\phi$, however, we do not observe a large difference between (i) and (ii).
In the analyses hereafter, we use the resummed $\eta_\phi$.

Including $\eta_\phi$, we obtain the fixed point values and the critical exponents for different gauge fixing values as follows:
\begin{itemize}
\item $\tilde\alpha\to 0$, $\beta=1$
\begin{align}
&   \tilde M_{P*}^2 = 0.036381,&
&\tilde V_* =0.0036794,&
& \eta_{\phi *} = -0.62968,&
\end{align}
\begin{align}
&        \theta_{1,2}= 2.40726 \pm 2.30995i,&
&       \theta_3= -0.74809,&
&       \theta_4= -2.1184,&
\end{align}
\item $\tilde\alpha\to 0$, $\beta=-1$
\begin{align}
&       \tilde M_{P*}^2 = 0.020658,&
&\tilde V_* =0.0023414,&
& \eta_{\phi*} = 1.4819,&
\end{align}
\begin{align}
&        \theta_1= 3.7098,&
&       \theta_2= 2.0541
&       \theta_3= 0.1844,&
&       \theta_4= -3.2975,&
\end{align}
\item $\tilde\alpha\to 0$, $\beta\to\infty$
\begin{align}
&       \tilde M_{P*}^2 = 0.019837,&
&\tilde V_* =0.00195508,&
& \eta_{\phi*} = 3.18076,&
\end{align}
\begin{align}
&        \theta_1= 4,&
&       \theta_2= 2.0795
&       \theta_3= -1.1808,&
&       \theta_4= -6.3615,&
\end{align}
\end{itemize}

In terms of the redeifned Planck mass, we have
\begin{itemize}
\item $\tilde\alpha\to 0$, $\beta=1$
\begin{align}
&   \widehat M_{P*}^2 = 0.012127,&
&\tilde V_* =0.0036794,&
& \eta_{\phi *} = 1.0067,&
\end{align}
\item $\tilde\alpha\to 0$, $\beta=-1$
\begin{align}
&       \widehat M_{P*}^2 =0.027544,&
&\tilde V_* =0.0023414,&
& \eta_{\phi*} = 1.0852,&
\end{align}
\end{itemize}
for which we obtain the same critical exponents as the original Planck mass.
The choice $\beta\to \infty$ does not yield the fixed point.

We find that the fixed point values are similar to the case with $\eta_\phi=0$, while the critical exponents of the scalar mass parameter and the quartic coupling change.
This is because an anomalous dimension of the scalar field of order one is induced. The beta functions for $\tilde M_P^2$ and $\tilde V$ have suppression factors, $\eta_\phi/4$ and $\eta_\phi/6$, as mentioned above, whereas the critical exponents of $\tilde m_H^2$ and $\tilde\lambda$ contains linear term in $\eta_\phi$ as given in Eqs.~\eqref{eq: approximated critical exponent of scalar mass} and \eqref{eq: approximated critical exponent of quartic}.
In particular, the choice $\beta\to \infty$ for the original Planck mass  yields $\eta_{\phi*} \approx 3$ which may be too large to accept the result.
We study the gauge dependence of $\eta_\phi$ in the next section and investigate a suitable choice for the gauge parameters.

\subsection{Anomalous dimensions as functions of the cosmological constant}

\begin{figure}
\begin{center}
\includegraphics[width=120mm]{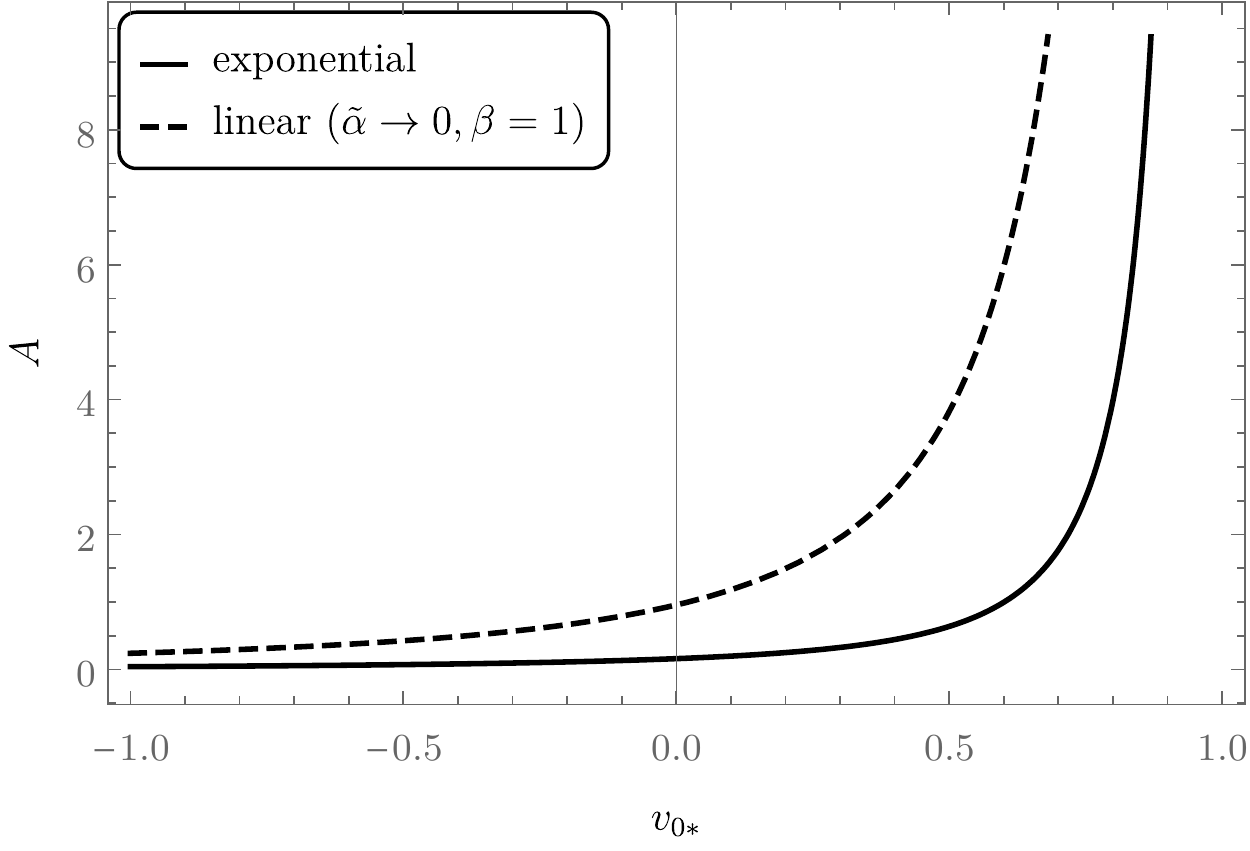}
\end{center}
\caption{
The metric-induced anomalous dimension $A$ defined in Eq.~\eqref{metric induced anomalous dimension in exp. par.} at $\widehat M_{P*}^2=1/8\pi$ as functions of $v_{0*}$ (solid line).
The dashed line shows Eq.~\eqref{eq: A in linear parametrization with alpha 0 and beta 1} case where $\tilde M_{P*}^2=1/8\pi$, the linear parametrization is used and the gauge parameters are chosen as $\tilde \a=0$, $\b=1$.}
\label{fig:critcosm1}
\end{figure}

Here we briefly examine how the critical exponents change as functions of the cosmological constant.
In particular, it was observed in the previous subsection and in \cite{Eichhorn:2017als} that the critical exponent for the Higgs mass changes sign when $\tilde V_*$ approaches the pole of the propagator. 
This casts some doubt on the validity of this behavior.
We have examined this and show the behavior of the metric-induced anomalous dimension $A$ in Fig.~\ref{fig:critcosm1}.
We plot $A$ for the exponential parametrization and the linear one with $\tilde\alpha\to 0$, $\beta=1$ for which $\tilde m_L^2=v_0$. We find that the metric-induced anomalous dimension in the linear parametrization takes a similar form to the case of the exponential parametrization:
\begin{align}
A\big|_\text{linear}&= \frac{3}{8\pi^2 \tilde M_{P*}^2}\frac{1}{(1-v_{0*})^2}.
\label{eq: A in linear parametrization with alpha 0 and beta 1}
\end{align}
For both linear and exponential parametrizations, there is a pole at $v_0=1$. In the linear parametrization, the TT mode in the metric fluctuations couples to the scalar potential and thus generates a finite contribution to $A$.
Hence, $A\big|_\text{linear}$ tends to be larger than $A\big|_\text{exp}$.

This is opposite to the  situation in the anomalous dimension of the scalar field.
As discussed in the beginning of this section, the TT mode in the linear parametrization contributes to $A$, but not to $\eta_\phi$.
On the other hand, in the exponential parametrization, $A$ does not receive the TT mode effects, whereas $\eta_\phi$ does.

\section{Gauge dependence}
\label{sec: gauge dependence}

Because some results depend on the choice of the gauge parameters, we would like to check quantitatively how the results change depending on the gauge parameters, and what physical implication may be extracted from the results even if they depend on the gauge.
Here we examine this by changing the gauge parameters in the exponential parametrization.
In terms of $\widehat M_P^2$ defined in Eq.~\eqref{eq: repacement explict}, the beta function of the cosmological constant has no apparent gauge dependence [see \eqref{eq: mass of scalar modes} and \eqref{eq: beta function of cosmological constant}]. In this section, we concentrate on that of fixed point values of the Planck mass and the anomalous dimension of the scalar field $\eta_\phi$.

\subsection{Planck mass and cosmological constant}

Let us look for fixed points of the Planck mass for various gauge parameters. To this end, we use the UV fixed point of the cosmological constant $\tilde V_*|_\text{UV}$ found in Eq.~\eqref{eq: fixed point of Lambda}.

We first recall the singularities in gauge parameters.
There is a singularity at $\b=3$. This singular behavior is observed in \cite{Gies:2015tca,Ohta:2016npm,Ohta:2016jvw}
and this is an artifact of our choice of gauge fixing function as discussed in Section~\ref{sec: Hessian}.
Besides, there is a singularity at $\tilde\alpha=3$: The flow equation for the original Planck mass squared \eqref{eq: flow equation of the Planck mass squared} has no singularity for any choice of $\tilde\alpha$, whereas the singularity at $\tilde\alpha=3$ does not cancel in the beta function of the redefined one.
This arises from the fact that the choice $\tilde\alpha=3$ erases the propagator of the trace mode $h$ which is the only field in the metric fluctuations directly coupled to the scalar potential with the exponential parametrization. See Eq.~\eqref{V1 explicit} in Appendix~\ref{app:sec:variation} and Eq.~\eqref{app: 2times2 propagator} in Appendix~\ref{app: sec: flow equation}.

\begin{figure}
\begin{center}
\includegraphics[width=66mm]{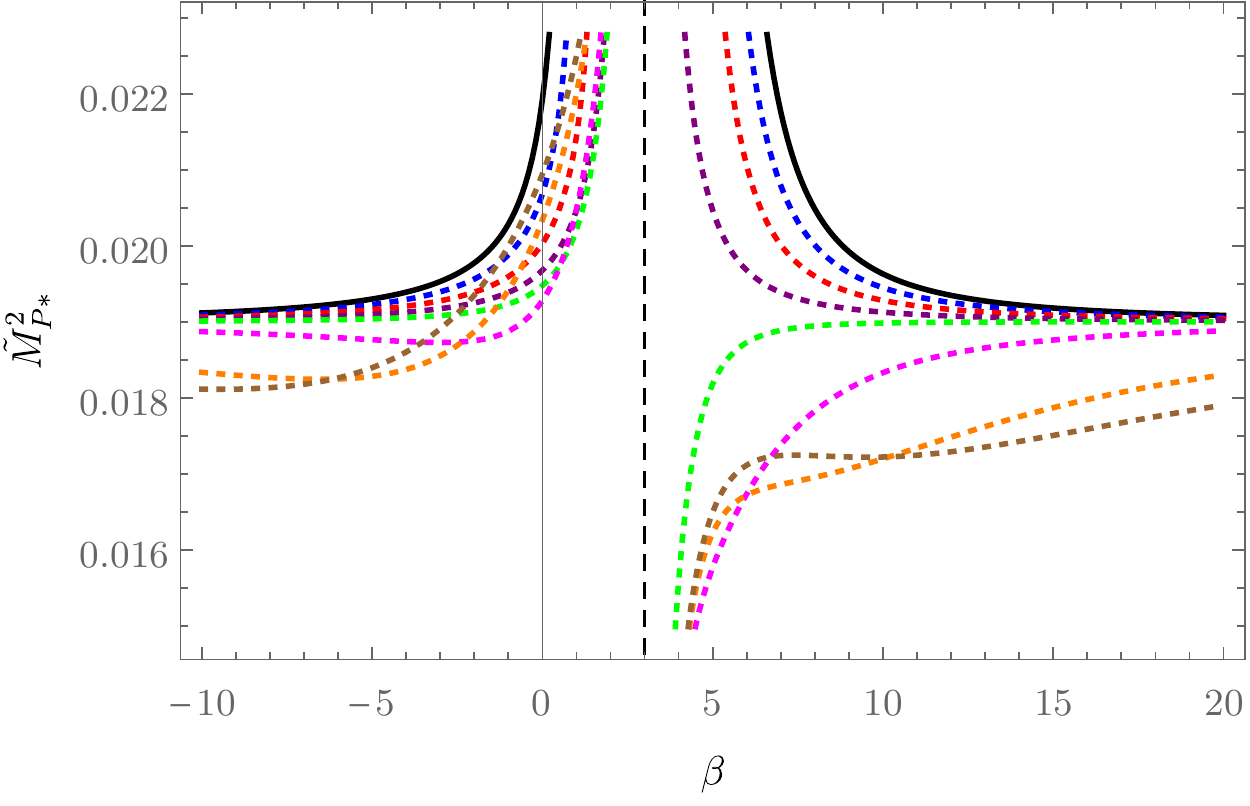}
\hs{2}
\includegraphics[width=89mm]{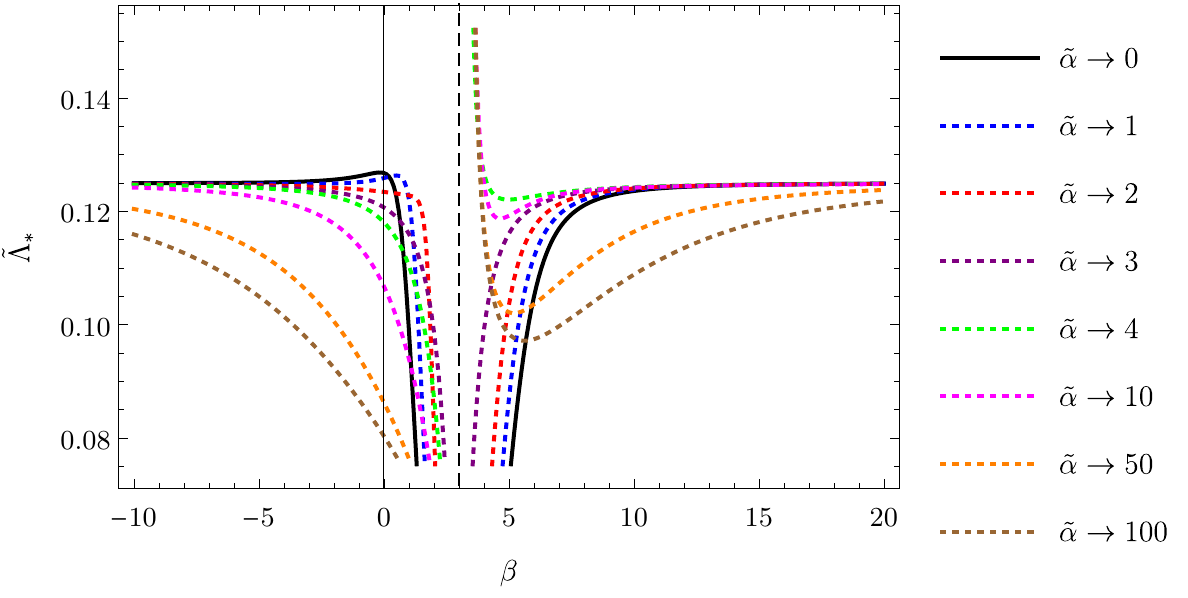}
\end{center}
\caption{Gauge dependence of the fixed point value of the Planck mass squared (left) the cosmological constant $\tilde \Lambda_*=\tilde V_*/\tilde M_{P*}^2$ (right). The vertical dashed line indicates the location of the pole at $\beta=3$.}
\label{fig:Mp_Lambda_original}
\end{figure}

We show the gauge dependence of the fixed point value of the Planck mass squared $\tilde M_{P}^2$ and the cosmological constant $\tilde \Lambda$ in Fig.~\ref{fig:Mp_Lambda_original}.
Around the pole at $\beta=3$, the fixed point values change drastically.
For $\beta$ well away from that pole, the fixed point values for any choice of $\tilde\alpha$ converge to the value $\tilde M_{P*}^2\approx 0.0190$ and $\tilde \Lambda_* \approx 0.125$ which are obtained by taking $\beta\to \pm \infty$. In the Landau gauge $\tilde\alpha\to 0$, such a convergence takes place faster.

\begin{figure}
\begin{center}
\includegraphics[width=66mm]{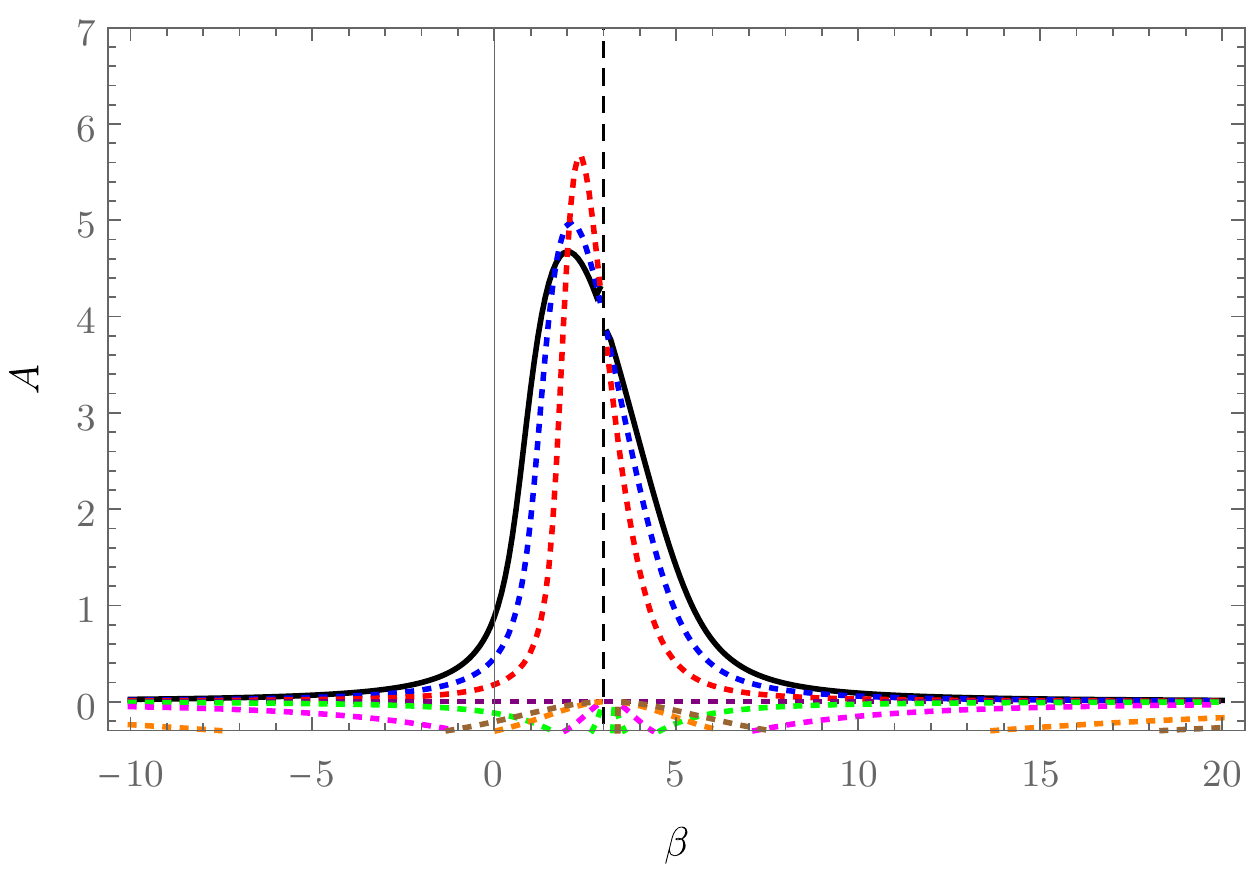}
\hs{2}
\includegraphics[width=89mm]{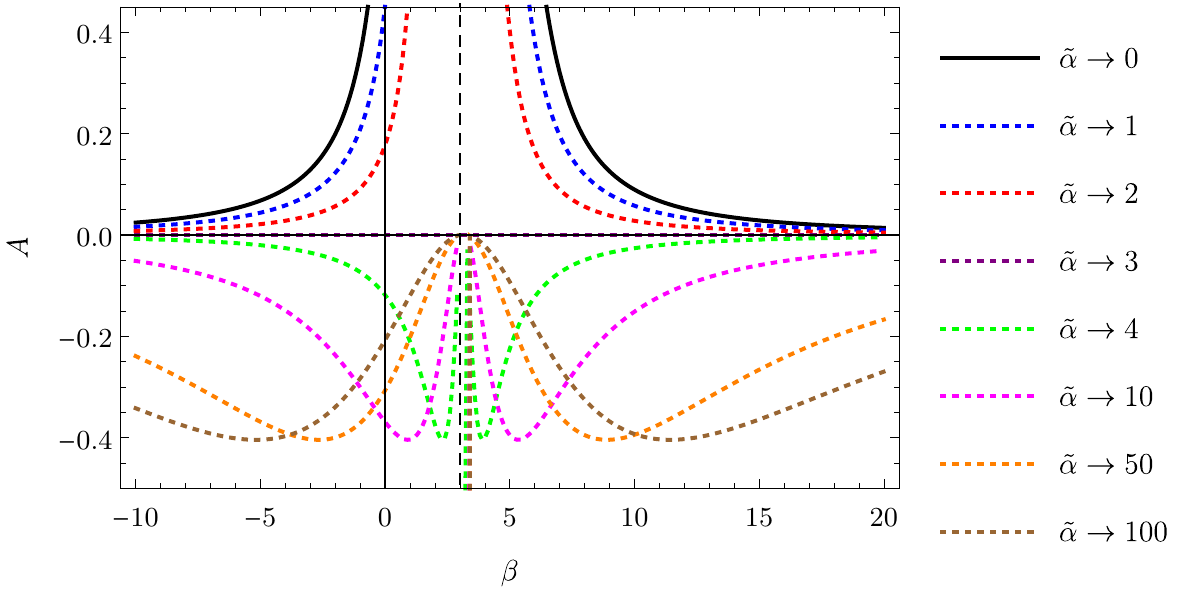}
\end{center}
\caption{Gauge dependence of the metric-induced anomalous dimension $A$ with $\tilde M_{P*}^2$ and $\tilde V_*$. The rhs panel shows the region of smaller $A$ in the lhs one.
The vertical dashed line locates at the pole of $\beta=3$.
}
\label{fig:gauge_dependence: A}
\end{figure}

In Fig.~\ref{fig:gauge_dependence: A}, we show the gauge dependence of the metric-induced anomalous dimension $A$ which is proportional to $(3-\tilde\alpha)$. Therefore, $A$ vanishes at $\tilde\alpha=3$ and flips the overall sign.
For $\tilde\alpha<3$, the metric-induced anomalous dimension exceeds 4 for $1<\beta<4$ and 2 for $0<\beta<5$, and these regions are not reliable

\begin{figure}[t]
\begin{center}
\includegraphics[width=150mm]{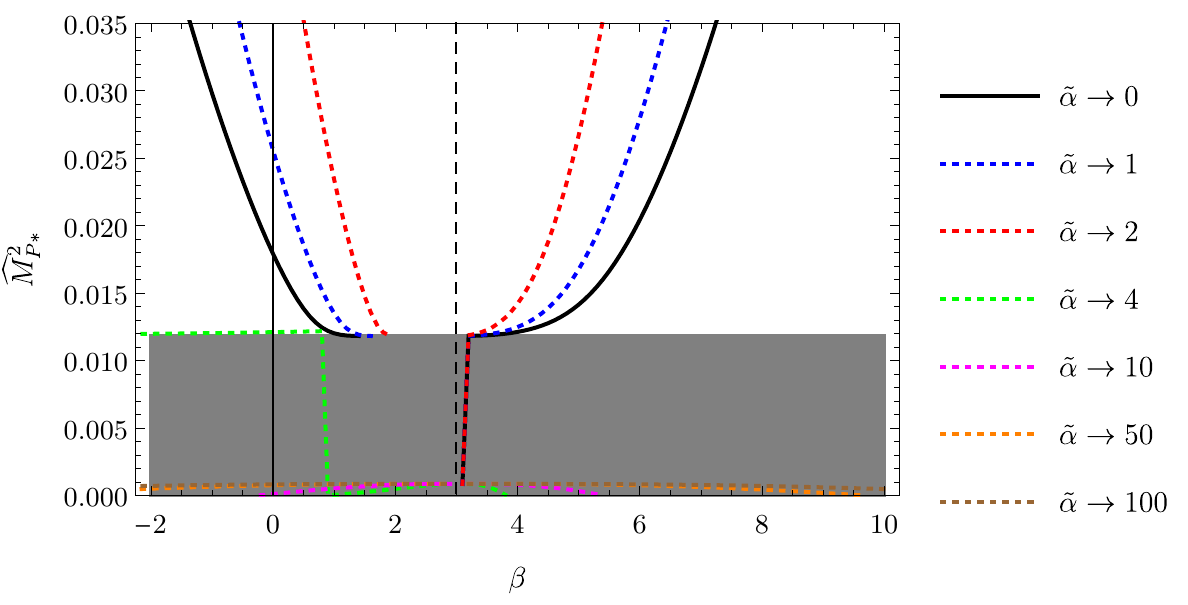}
\end{center}
\caption{Gauge dependence of the redefined Planck mass squared $\widehat M_{P*}^2$.
In the gray region, no UV fixed point value of $\tilde V$ is found.
The vertical dashed line corresponds to the pole at $\beta=3$.
}
\label{fig:pm}
\end{figure}

\begin{figure}[htb]
\begin{center}
\includegraphics[width=150mm]{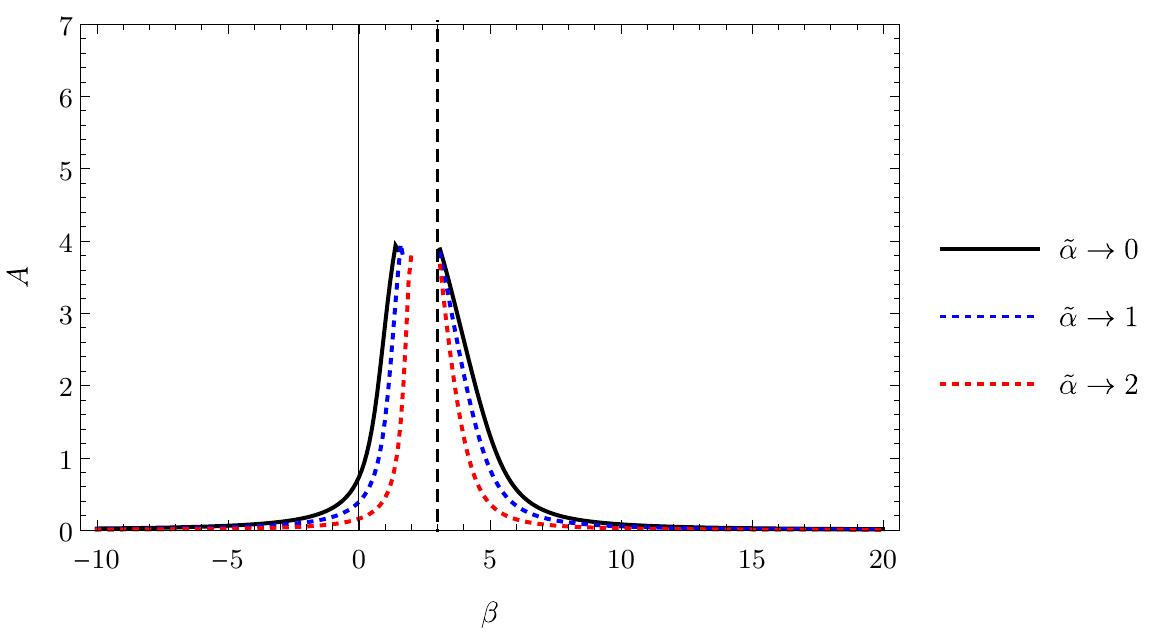}\\[3ex]
\end{center}
\caption{Gauge dependence of the metric-induced anomalous dimension $A$ with $\widehat M_{P*}^2$ and $\tilde V_*|_\text{UV}$.
The pole at $\beta=3$ is shown by the vertical dashed line.
}
\label{fig:gauge_dependence: A redefined}
\end{figure}

Let us next check the redefined Planck mass.
The gauge dependence of $\widehat M_{P*}^2$ is depicted in Fig.~\ref{fig:pm}. Here we have used the fixed point $\tilde V_*|_\text{UV}$ given in Eq.~\eqref{eq: fixed point of Lambda} and therefore there is no reliable fixed point of $\tilde V$ for $\widehat M_P^2 \leq\frac{2+\sqrt{3}}{32\pi^2}\simeq 0.01182$. 
This region is shown by gray in Fig.~\ref{fig:pm}.
We see from this figure that $\widehat M_P^2$ has a strong gauge dependence which is imported from the mass of scalar modes $\tilde m_s^2$.
For $1<\beta<4$ in $\tilde\alpha<3$, and for any $\beta$ in $\tilde\alpha>3$, the fixed point value of $\widehat M_{P}^2$ is not found. Those just correspond to the region where $A$ exceeds 4 or becomes negative. Thus, we should not take these gauge fixing parameters; the preferable region is $\tilde\alpha<3$ and $\beta<1$.
Note that for $\beta\to \infty$, we have $\widehat M_{P*}^2$ to be infinity. This is consistent with the fact that the metric fluctuation decouples from the scalar potential in the limit $\beta\to \infty$ as we have seen in the previous section.
The gauge dependence of the metric-induced anomalous dimension is shown in Fig.~\ref{fig:gauge_dependence: A redefined}. This is the same as in Fig.~\ref{fig:gauge_dependence: A} except for $1<\beta<4$ in which $A$ exceeds $4$.


From the above analysis, we conclude that the region where the metric-induced anomalous dimension is larger than $4$ should be excluded. This means that the cosmological constant cannot be irrelevant within the local potential approximation with $\eta_\phi=0$.

Even if $A>4$ is admitted, the cosmological constant cannot be irrelevant. 
To see this, we examine more precise behavior of the critical exponents of $\tilde M_P^2$ and $\tilde V$ given by the eigenvalues
of the stability matrix \eqref{eq: stabilit matrix of PM and CC}: 
\begin{align}
\theta_{\tilde M_P^2, \tilde V} 
= -\frac{1}{2}\left.\left\{ \frac{\p \beta_{\tilde M_P^2}}{\p \tilde M_P^2} + \frac{\p \beta_{\tilde V}}{\p \tilde V}  \pm \left(\frac{\p \beta_{\tilde M_P^2}}{\p \tilde M_P^2} - \frac{\p \beta_{\tilde V}}{\p \tilde V} \right)\sqrt{1 + \frac{4\frac{\p \beta_{\tilde V}}{\p \tilde M_P^2} \frac{\p \beta_{\tilde M_P^2}}{\p \tilde V}}{\left( \frac{\p \beta_{\tilde M_P^2}}{\p \tilde M_P^2} - \frac{\p \beta_{\tilde V}}{\p \tilde V} \right)^2}}  \right\}\right|_{\substack{\tilde M_P^2=\tilde M_{P*}^2\\ \tilde V= \tilde V_{*}}}.
\label{eq: eigenvalues of stability matrix}
\end{align}
When the off-diagonal parts are negligible, the critical exponents are simply given by the diagonal parts:
\begin{align}
&     \theta_{\tilde M_P^2}\approx -\frac{\p \beta_{\tilde M_P^2}}{\p \tilde M_P^2},&
&      \theta_{\tilde V} \approx -\frac{\p \beta_{\tilde V}}{\p \tilde V}=4-A.
\end{align}
On the other hand, when the off-diagonal parts are large so that inside of the square root in Eq.~\eqref{eq: eigenvalues of stability matrix} becomes negative, the eigenvalues in Eq.~\eqref{eq: eigenvalues of stability matrix} become imaginary and thus the real part of the critical exponents is commonly given by 
\begin{align}
\text{Re}\,[\theta_{\tilde M_P^2, \tilde V} ]\approx -\frac{1}{2}\left( \frac{\p \beta_{\tilde M_P^2}}{\p \tilde M_P^2} + \frac{\p \beta_{\tilde V}}{\p \tilde V}\right)\Bigg|_{\substack{\tilde M_P^2=\tilde M_{P*}^2\\ \tilde V= \tilde V_{*}}},
\label{eq: approximated critical exponents}
\end{align}
for both $\tilde M_{P*}^2$ and $\tilde V_{*}$.
This is what happens in Eq.~\eqref{eq: critical exponents in orgial Mp in alpha0 beta1}.

\begin{figure}
\begin{center}
\includegraphics[width=150mm]{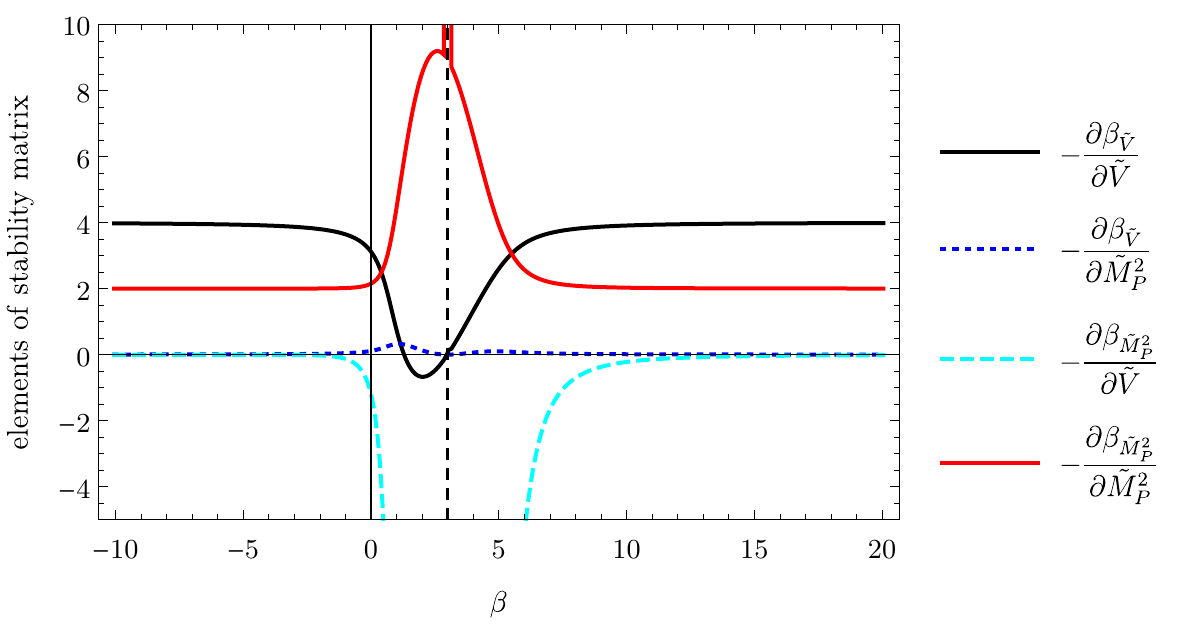}
\end{center}
\caption{
Gauge dependence of each element of the stability matrix \eqref{eq: stabilit matrix of PM and CC} in the Landau gauge $\tilde\alpha\to 0$.
The position of the pole at $\beta=3$ is shown by the vertical dashed line.
}
\label{fig:elements_stability}
\end{figure}

\begin{figure}
\begin{center}
\includegraphics[width=77mm]{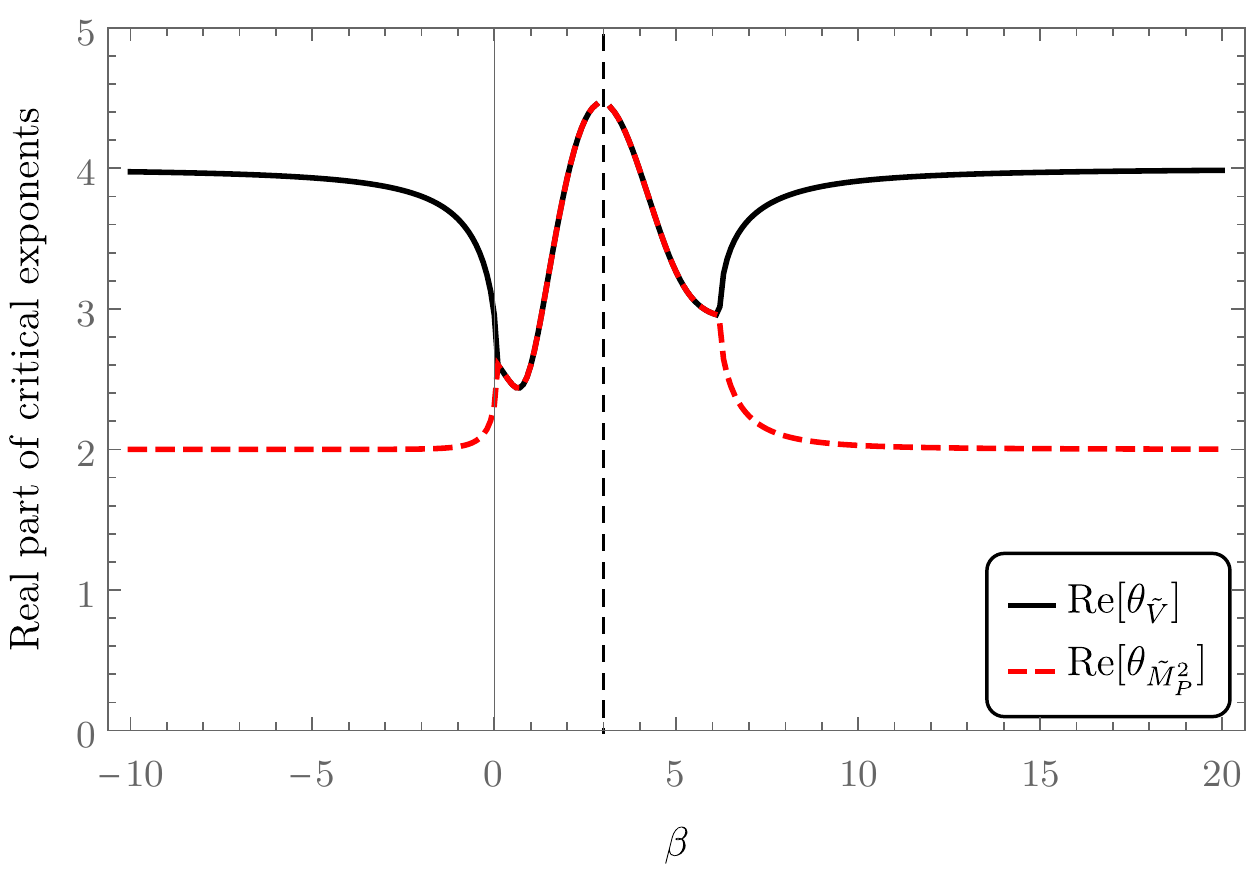}
\hs{2}
\includegraphics[width=77mm]{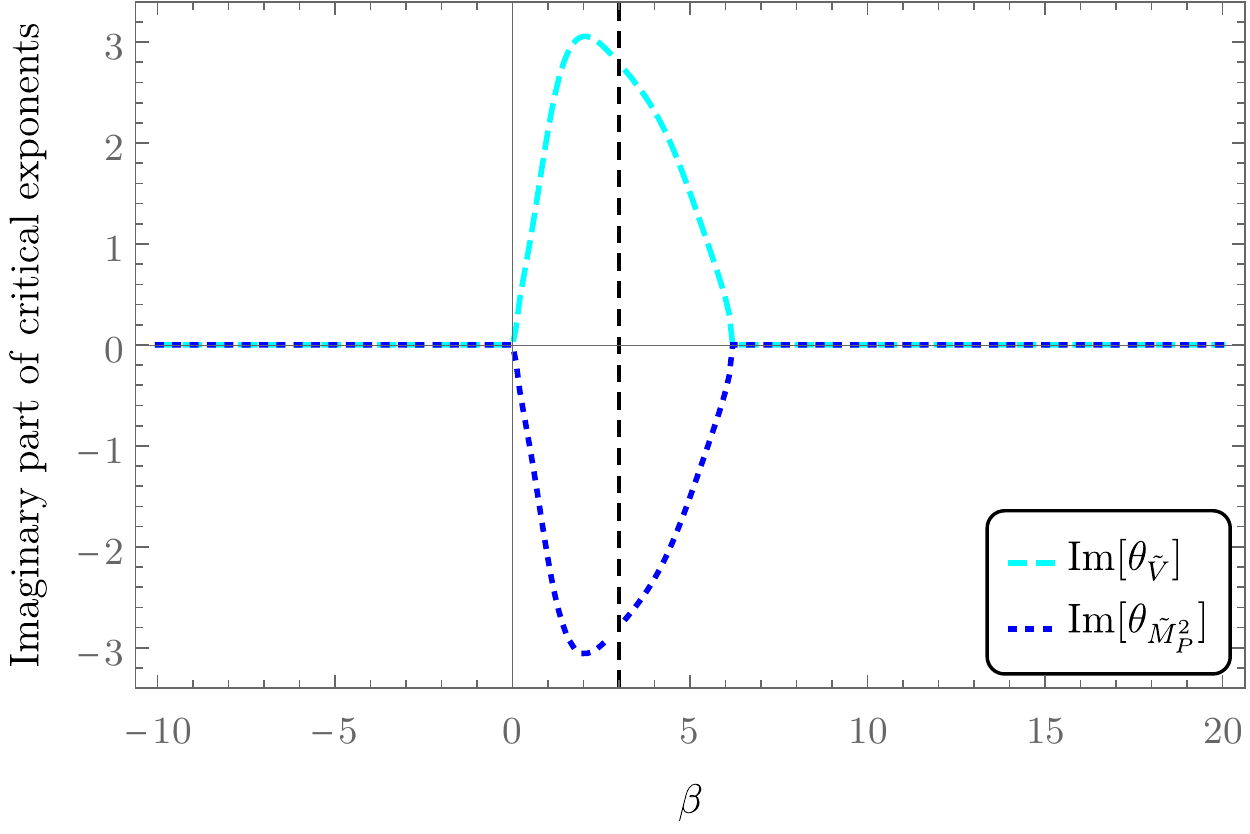}
\end{center}
\caption{Gauge dependence of the real (left) and imaginary (right) parts of the critical exponents $\theta_{\tilde V}$, $\theta_{\tilde M_P^2}$ in the Landau gauge $\tilde \alpha\to 0$.
The vertical dashed line shows the location of the pole at $\beta=3$.
}
\label{fig:gauge_dependence: critica exponents of M and V}
\end{figure}

In Fig.~\ref{fig:elements_stability}, we plot the $\beta$ dependence of each element of the stability matrix in the Landau gauge $\tilde\alpha\to0$.
For $\beta$ well away from the singularity at $\beta=3$, the off-diagonal parts are negligibly small. In the region $0<\beta<6$, however, the off-diagonal part $\frac{\p \beta_{\tilde M_P^2}}{\p \tilde V}$ is significantly large, so that the critical exponents tend to be imaginary. Moreover, we see for $1<\beta<4$ that $-\frac{\p \beta_{\tilde V}}{\p \tilde V}$ turns to negative due to $A>4$, while $-\frac{\p \beta_{\tilde M_P^2}}{\p \tilde M_P^2}$ is positive and large.
Hence, the real part of the critical exponents \eqref{eq: approximated critical exponents} remains positive and the cosmological constant does not become irrelevant at least in the current setup.
See Fig.~\ref{fig:gauge_dependence: critica exponents of M and V} in which the $\beta$-dependence of the real and imaginary parts of the critical exponents, $\theta_{\tilde M_P^2,\tilde V}$, are shown for the Landau gauge $\tilde \alpha\to0$. One can see that the imaginary part appears for $0<\beta<6$.

Figs.~\ref{fig:elements_stability} and \ref{fig:gauge_dependence: critica exponents of M and V} also tell us that the choice around $0<\beta<6$ may not be appropriate in the sense of the principle of minimal sensibility.
We thus conclude again that the values of $\beta$ in this region which make $-\frac{\p \beta_{\tilde V}}{\p \tilde V}=4-A$ negative should be excluded if the redefined Planck mass is used.

\subsection{Anomalous dimension of the scalar field}

Here we study the gauge dependence of the anomalous dimension of the scalar field $\eta_\phi$.
Whereas we would have concluded that the gauge choice $0<\beta<5$ should be avoided, we have to pay attention to the magnitude of $\eta_\phi$.
A large value of $\eta_\phi$ violates the validity of the derivative expansion for the effective action of the scalar field. 
In particular, we expect that in the exponential parametrization, the tadpole diagram with a loop of the TT mode gives a significant contribution to $\eta_\phi$. See Eq.~\eqref{eq: diagrammatic flow equation of Zphi in exponential}.
The explicit form of $\eta_\phi$ is given in Eq.~\eqref{anomalous dimension etaphi}.
Indeed, in Section~\ref{sec: Critical exponent of the scalar effective potential}, we have observed large values of $\eta_\phi$ for certain gauge parameter choices, e.g. $\eta_\phi \approx 3$ for $\beta\to \infty$ and $\tilde\alpha\to 0$.

\begin{figure}
\begin{center}
\includegraphics[width=150mm]{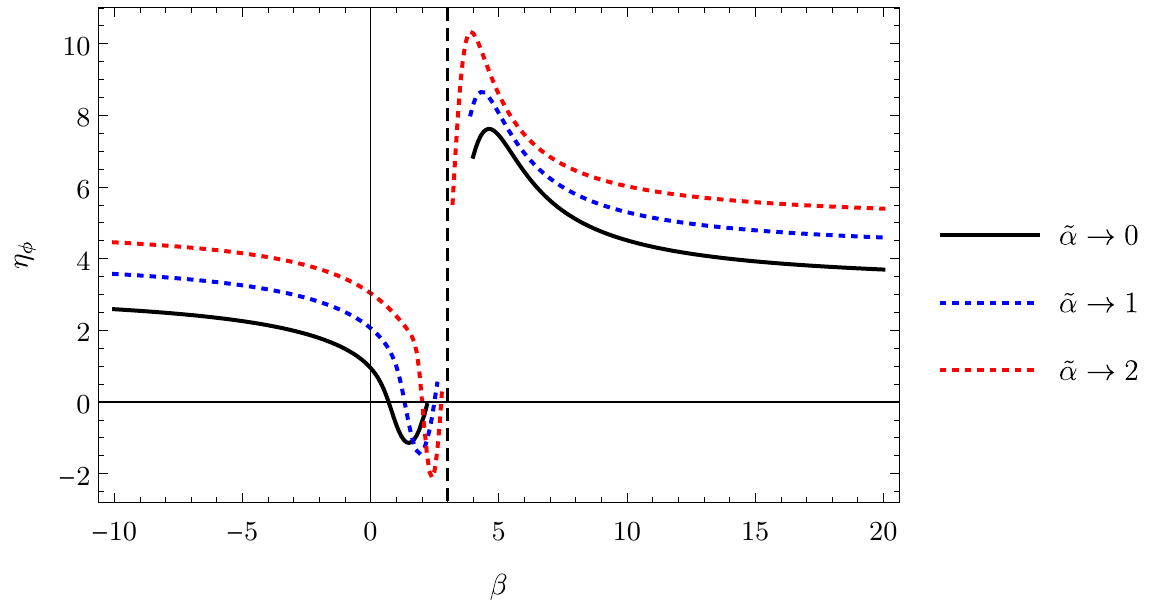}
\end{center}
\caption{
Gauge dependence of $\eta_\phi$.
The vertical dashed line is located at $\beta=3$.
}
\label{fig:gauge dependence of eta_phi}
\end{figure}

We show the behavior of $\eta_\phi$ in \eqref{anomalous dimension etaphi} as a function of $\beta$ for $\tilde\alpha=0,\,1,\,2$ in Fig.~\ref{fig:gauge dependence of eta_phi}.
First, we see that $\eta_\phi$ takes larger values if $\beta>3$ for any $\tilde\alpha$. This is not acceptable in terms of the validity of the derivative expansion.
Second, although $\eta_\phi$ for $0<\beta<3$ is relatively smaller, it is sensitive to the variation in $\beta$. For $\beta<0$ and $\tilde\a \neq 0$, one has $\eta_\phi\approx 3$ or larger which may be out of the validity of the derivative expansion.
If we permit $\eta_\phi\approx 1$, the choice $-5<\beta<0$ with $\tilde\a=0$ seems to be suitable together with the principle of minimal sensitivity.

\section{Summary and Conclusions}
\label{sec: conclusions}

We have studied the parametrization and gauge dependences in the Higgs-gravity system by using the functional renormalization group.
We have considered the Einstein-Hilbert truncation for a gravitational sector and used the exponential parametrization and compared the results in the linear parametrization.
Assuming an Einstein spacetime for the background field, the beta functions for the Planck mass squared, the cosmological constant and the scalar potential have been derived.
We have used these results to study the parametrization and gauge dependences of the fixed points and the critical exponents.
We find that the results do depend on these, but if we choose suitable range of gauge parameters, we may get reasonable and stable results under the change of gauge parameters, in accord with the principle of minimum sensitivity.

For the comparison of the parametrization dependence, the main features we have found are the following:
\begin{enumerate}
\item 
An advantage of the exponential parametrization is that only the trace mode $h$ in the metric fluctuations couples to the scalar potential including the cosmological constant. Consequently, the propagators of the other modes ($h_{\mu\nu}^{TT}$, $\xi_\mu$ and $\sigma$) in metric fluctuations have no artefactual poles arising from the cosmological constant.
\item
The gauge parameters appear in the scalar potential in a specific combination with the Planck mass, and thus the beta function of the scalar potential is written
in an apparently gauge-independent form by the redefinition of the Planck mass~\eqref{eq: repacement explict}.
This allowed us to discuss the beta functions without gauge dependence.
\item
These advantages, however, are valid only if we can use the local potential approximation for the scalar field, i.e. $Z_\phi=1$ (or equivalently $\eta_\phi=0$). In the exponential parametrization,
we have found that the loop corrections to $\eta_\phi$ come from all modes including the TT mode in the metric fluctuations which entail $\eta_\phi$ to be large.
In the pure gravity, it appears that large $|\b|$ is favored, but when the scalar fields exist, the gravity decouples from the scalar fields in the limit, and gives large contribution to the anomalous dimensions.
Thus the exponential parametrization has also its limitation, and the linear parametrization may be better in this respect.

\item
These situations are opposite in the linear parametrization:
the scalar potential receives loop effects of all modes in metric fluctuations, whereas only the spin-1 transverse vector and the scalar modes contribute to $\eta_\phi$.
Hence, although $\eta_\phi$ is small, the scalar potential have loop contributions from all modes which are strongly gauge dependent.
\end{enumerate}

For the gauge dependence of the fixed points and critical exponents, we have shown
\begin{enumerate}
\item 
There are singularities at $\b=3$ and $\tilde\a=3$, and we should be well away from these singularities to get stable results.
It appears that $\tilde\a<3$ is preferable, and $\tilde\a=0$ gives reasonable result. This point is also expected to be a fixed point if we let $\tilde\a$ run under the renormalization group.

\item
In the pure gravity, it appears that large $|\b|$ is favored. When the scalar fields exist, the gravity decouples from the scalar fields in the limit, but gives large contribution to the anomalous dimensions. Considering the result on the gauge dependence of the anomalous dimensions, we find that $-5\leq \b\leq -1$ and $\tilde\a=0$ gives reasonable results.

\end{enumerate}

These investigations imply that the optimization in the gravitational sector tends to be in conflict with that in the Higgs sector.
One needs to find a suitable choice for parameters such that both gravity and matter sectors are reasonably optimized.
In our study, we have found that the region $\tilde\a\sim 0, -5\leq\b\leq -1$ seems to give optimized choice and the critical exponents are around
\bea
\t_1\sim 3.7, \qquad
\t_2 \sim 2.1, \qquad
\t_3 \sim 0.2, \qquad
\t_4 \sim -3,
\eea
where $\eta_\phi$ gives relatively larger effect than $A$ on the scalar mass parameter and the quartic coupling.
Although the critical exponent of the Higgs mass parameter tends to be smaller than the canonical one, it still remains to be a relevant coupling within the present setups.
We thus conclude that the critical exponent for the quartic coupling is negative, so that it is irrelevant, leading to the prediction to the low-energy physics, but other couplings are likely to be relevant.
In particular we have shown that the cosmological constant never becomes irrelevant.

In general, it is expected that the improvement of truncation makes parameter dependence mild. For instance, it has been actually seen in Ref.~\cite{Aoki:2012mj} that by taking sub-leading effects into account, one can realize almost the gauge-parameter independence of the order parameter for spontaneous chiral symmetry breaking in quantum chromodynamics.
An interesting question therefore is how much gauge dependence is reduced by the improvement of the gravitational sector, especially by constructing the full non-perturbative propagator~\cite{Bosma:2019aiu,Knorr:2019atm,Bonanno:2021squ,Knorr:2021niv}.
It is also expected that systematic improvements of vertices~\cite{Christiansen:2012rx,Christiansen:2014raa,Christiansen:2015rva,Meibohm:2016mkp,Christiansen:2016sjn,Denz:2016qks,Christiansen:2017cxa,Christiansen:2017bsy,Eichhorn:2018akn,Eichhorn:2017als} beyond the background-field approximation reduce gauge and parametrization dependences.

\subsubsection*{Acknowledgment}

We thank Roberto Percacci for valuable discussions.
The work of N.\,O. was supported in part by the Grant-in-Aid for Scientific Research Fund of the JSPS (C) No.~16K05331, No.~20K03980, and Taiwan MOST 110-2811-M-008-510.
The work of M.\,Y. is supported by the Alexander von Humboldt Foundation, the DFG Collaborative Research Centre ``SFB 1225 (ISOQUANT)", and Germany’s
Excellence Strategy EXC-2181/1-390900948 (the Heidelberg Excellence Cluster STRUCTURES).

\appendix

\section{York decomposition}
\label{york}

The York decomposition is defined by
\bea
h_{\mu\nu} = h^{TT}_{\mu\nu} + \nabla_\mu\xi_\nu +\nabla_\nu\xi_\mu +
\left(\nabla_\mu \nabla_\nu -\frac{1}{4} \bg_{\mu\nu} \nabla^2\right) \s +
\frac{1}{4} \bg_{\mu\nu} h,
\label{eq:york}
\eea
where
\bea
\nabla^\mu h^{TT}_{\mu\nu} = \bg^{\mu\nu} h^{TT}_{\mu\nu}
= \nabla_\mu \xi^\mu=0.
\eea
When Eq.~\eqref{eq:york} is squared, we get
\bea
\int d^4 x\sqrt{\bar g}\Big[ h^{TT}_{\mu\nu}h^{TT\,\mu\nu} +2 \xi_\mu \left(\lich_{1}
-\frac{1}{2}\br \right)\xi^\mu +\frac{3}{4}\s \lich_{0}\left( \lich_{0}-\frac{\br}{3} \right) \s
+\frac{1}{4} h^2 \Big],
\label{app: squared metric fluctuation}
\eea
where $\Delta_{Li}$ are Lichnerowicz Laplacians defined in Appendix~\ref{lich}.
Note that we can freely insert the covariant derivatives inside the above expression.

Let us here consider a Gaussian integral
\begin{align}
1=\int {\mathcal D}h_{\mu\nu} \,e^{-\int d^4 x\sqrt{\bg}\, h^{\mu\nu} h_{\mu\nu}}.
\end{align}
Due to the York decomposition, this integral is rewritten as
\begin{align}
1= J\int {\mathcal D}h^{TT}_{\mu\nu}{\mathcal D}\xi_{\mu} {\mathcal D}\sigma {\mathcal D}h  \,e^{-\int d^4 x\sqrt{\bg}\left[  h^{TT}_{\mu\nu}h^{TT\,\mu\nu} +2 \xi_\mu \left(\lich_{1}
-\frac{1}{2}\br \right)\xi^\mu +\frac{3}{4}\s \lich_{0}\left( \lich_{0}-\frac{\br}{3} \right) \s
+\frac{1}{4} h^2\right] }.
\end{align}
in which by performing the Gaussian integrals for each fields, the Jacobian should be identified with
\begin{align}
J =\Det_{(1)}\left(\lich_1-\frac{\br}{2}\right)^{1/2}
\Det_{(0)}\left[\lich_0 \left(\lich_0-\frac{\br}{3}\right)\right]^{1/2}
\equiv J_\text{grav1} J_\text{grav0},
\end{align}
where the suffices indicate which spin contributes. If we use new field variables
\bea
\hat\xi_\mu=\sqrt{\lich_1-\frac{\br}{2}}\ \xi_\mu;\qquad
\hat\s =\sqrt{\lich_0} \sqrt{\lich_0-\frac{\br}{3}}\  \s,
\eea
we do not have to introduce the Jacobian.

A similar decomposition should be made for the ghost fields which is given in Eq.~\eqref{eq: decomposition of ghost field}. Thanks to the inverse of the squared (Lichnerowicz) Laplacian in front of the scalar ghost $C^L$, the Jacobian associated to the decomposition does not arise.

\section{Lichnerowicz Laplacians}
\label{lich}

The Lichnerowicz Laplacians are defined as
\bea
\lich_{2} T_{\mu\nu} &=& -\nabla^2 T_{\mu\nu} +R_\mu{}^\rho T_{\rho\nu}
+ R_\nu{}^\rho T_{\mu\rho} -R_{\mu\rho\nu\s} T^{\rho\s} -R_{\mu\rho\nu\s} T^{\s\rho}, \nn
\lich_{1} V_\mu &=& -\nabla^2 V_\mu + R_\mu{}^\rho V_\rho, \nn
\lich_{0} S &=& -\nabla^2 S.
\eea
These operators have the useful properties of ``commuting with covariant derivative''
in the sense that
\bea
\lich_{2} (\nabla_\mu \xi_\nu + \nabla_\nu \xi_\mu)
&=& \nabla_\mu \lich_{1} \xi_\nu + \nabla_\nu \lich_{1} \xi_\mu, \nn
\lich_{2} (\nabla_\mu \nabla_\nu S)
&=& \nabla_\mu \nabla_\nu \lich_{0} S.
\eea

\section{Variations}
\label{app:sec:variation}

We give variations for the effective action
\begin{align}
\Gamma_k=\frac{Z_\phi}{2}\int d^4x\sqrt{g} g^{\mu\nu} \partial_\mu \phi \partial_\nu \phi
+ \int d^4x\sqrt{g}\left[ U(\rho) - Z_NR \right].
\end{align}
With $\phi=\bar\phi+\varphi$ and the exponential parametrization~\eqref{exp}, the second order of fluctuation fields in the effective action reads
\begin{align}
\delta^2 \Gamma_k
&= \frac{Z_\phi}{2} \int d^4x\sqrt{\bar g}\left[ 
H_2^{\mu\nu} \partial_\mu \bar\phi \partial_\nu \bar\phi
+H_1^{\mu\nu} \partial_\mu \varphi \partial_\nu \bar\phi
\right]+ \int d^4x\sqrt{\bar g}\left[ V_1 - Z_N V_2 \right]\,,
\end{align}
where
\begin{align}
H_1^{\mu\nu}&=  h \bar g^{\mu\nu}- 2h^{\mu\nu},
\label{eq: decomposed H1}
\\
H_2^{\mu\nu}&= \frac{1}{8}h^2 \bar g^{\mu\nu} +\frac{1}{2}h^\mu{}_\lambda h^{\nu\lambda}  - \frac{1}{2}h h^{\mu\nu},
\label{H2 explicit}
\\
V_1&= \frac{1}{2} \left( \frac{h^2}{4} U(\bar\rho) + \left(U'(\bar\rho)+ 2 U''(\bar\rho)\bar\rho \right) \varphi^2 +U'(\bar\rho)\sqrt{2\bar{\rho}}\,h \varphi  \right),
\label{V1 explicit}
\\
V_2&=\frac{1}{4}h^2 \bar R + h^{\mu\nu}\nabla_\mu \nabla_\nu h -2 h^{\mu\nu}\nabla_\nu \nabla_\alpha h_\mu{}^\alpha + h^{\mu\nu}\nabla^2 h_{\mu\nu} +\frac{3}{4}\nabla_\alpha h_{\mu\nu} \nabla^\alpha h^{\mu\nu} -\frac{1}{4}\nabla_\nu h \nabla^\nu h \nn
&\quad
-\nabla_\mu h^{\mu\nu} \nabla_\alpha h_\nu{}^\alpha
+\nabla^\nu h \nabla^\alpha h_{\nu\alpha} -\frac{1}{2}\nabla_\nu h_{\mu\alpha} \nabla^\alpha h^{\mu\nu} +\frac{1}{4}\nabla_\mu \nabla_\nu h^{\mu\alpha}h_{\alpha}{}^\nu -\frac{1}{4}\nabla^2 h^{\mu\alpha}h_{\alpha\mu} \nn
&\quad
+ h^{\mu\nu}h_{\alpha\beta}\bar R_{\mu\alpha\nu\beta} -\frac{1}{4}h^{\mu\alpha}h_{\alpha}{}^\nu \bar R_{\mu\nu}
+h (\nabla_\mu \nabla_\nu h^{\mu\nu} -\nabla^2 h -h^{\mu\nu}\bar R_{\mu\nu}).
\end{align}
Note that in this work, we make the replacement $h_{\mu\nu}\to Z_N^{-1/2}h_{\mu\nu}$.

After employing the York decomposition, the Hessian has the following structure:
\begin{align}
\Gamma_k^{(2)}= \frac{1}{2}\pmat{
(\Gamma_{h^{TT}h^{TT}})_{\mu\nu}{}^{\rho\sigma} & 0 & 0 & ({\mathcal V}_{h^{TT} h})_{\mu\nu} & ({\mathcal V}_{h^{TT} \varphi})_{\mu\nu} \\[2ex]
0 & (\Gamma_{\xi\xi})_\mu{}^\rho & 0 & 0 & ({\mathcal V}_{\xi\varphi})_{\mu} \\[2ex]
0 & 0 & \Gamma_{\sigma\sigma} & \Gamma_{\sigma h} & \Gamma_{\sigma\varphi}\\[2ex]
({\mathcal V}_{h h^{TT}})^{\rho\sigma} & 0 & \Gamma_{h\sigma} & \Gamma_{hh} & \Gamma_{h\varphi}\\[2ex]
({\mathcal V}_{\varphi h^{TT} })^{\rho\sigma} & ({\mathcal V}_{\varphi \xi})^{\rho} & \Gamma_{\varphi\sigma} & \Gamma_{\varphi h} & \Gamma_{\varphi\varphi}
},
\end{align}
where each component is separated into the kinetic term and the vertex as
\begin{align}
(\Gamma_{h^{TT}h^{TT}})_{\mu\nu}{}^{\rho\sigma} &= ({\mathcal K}_{h^{TT}h^{TT}})(P_{TT})_{\mu\nu}{}^{\rho\sigma} + ({\mathcal V}_{h^{TT}h^{TT}})_{\mu\nu}{}^{\rho\sigma}, \\[2ex]
(\Gamma_{\xi\xi})_\mu{}^\rho&=({\mathcal K}_{\xi\xi})(P_{1})_{\mu}{}^{\rho} + ({\mathcal V}_{\xi\xi})_{\mu}{}^{\rho}, \\[2ex]
\pmat{
\Gamma_{\sigma\sigma} & \Gamma_{\sigma h} & \Gamma_{\sigma\varphi}\\[2ex]
\Gamma_{h\sigma} & \Gamma_{hh} & \Gamma_{h\varphi}\\[2ex]
\Gamma_{\varphi\sigma} & \Gamma_{\varphi h} & \Gamma_{\varphi\varphi}
}
&= \pmat{
{\mathcal K}_{\sigma\sigma} & {\mathcal K}_{\sigma h} & 0\\[2ex]
{\mathcal K}_{h\sigma} & {\mathcal K}_{hh} & 0 \\[2ex]
0 & 0 & {\mathcal K}_{\varphi\varphi}
}
+
\pmat{
{\mathcal V}_{\sigma\sigma} & {\mathcal V}_{\sigma h} & {\mathcal V}_{\sigma\varphi}\\[2ex]
{\mathcal V}_{h\sigma} & {\mathcal V}_{hh} & {\mathcal V}_{h\varphi}\\[2ex]
{\mathcal V}_{\varphi\sigma} & {\mathcal V}_{\varphi h} & {\mathcal V}_{\varphi\varphi}
}.
\end{align}
Here, $(P_{TT})_{\mu\nu}{}^{\rho\sigma}$ is the transverse-traceless (TT) projector which is given in a flat spacetime background, where the Fourier transformation can be used, as 
\begin{align}
(P_{TT}(q))_{\mu\nu}{}^{\rho\sigma}
=\frac{1}{2}\Big[ (P_1(q))_\mu{}^{\rho}  (P_1(q))_\nu{}^{\sigma} +  (P_1(q))_\mu{}^{\sigma}  (P_1(q))_\nu{}^{\rho} \Big] - \frac{1}{3} (P_1(q))_{\mu\nu}  (P_1(q))^{\rho\sigma},
\label{app: TT projector}
\end{align}
with the transverse vector projector
\begin{align}
(P_1(q))_{\mu\nu} = \delta_{\mu\nu} -\frac{q_\mu q_\nu}{q^2}.
\end{align}
The kinetic terms $\mathcal K$ are given in Eqs.~\eqref{eq: Hessian for TT mode}, \eqref{eq: Hessian for xi} and \eqref{eq: Hessian for scalar fields} without the $\bar\phi$-dependence.
In particular, the kinetic term of the scalar modes is
\begin{align}
\pmat{
{\mathcal K}_{\sigma\sigma} & {\mathcal K}_{\sigma h} & 0\\[2ex]
{\mathcal K}_{h\sigma} & {\mathcal K}_{hh} & 0 \\[2ex]
0 & 0 & {\mathcal K}_{\varphi\varphi}
} &=\small{
\left(
\begin{array}{ccc}
-\frac{3}{16} (\lich_0)^2 \left(\lich_0-\frac{\br}{3}\right) &
-\frac{3}{16} \lich_0 \left(\lich_0-\frac{\br}{3}\right) &
0 \\[2ex]
-\frac{3}{16} \lich_0 \left(\lich_0-\frac{\br}{3}\right) &
-\frac{3}{16} \left(\lich_0+\frac{\br}{3}\right) +\frac{1}{4}\frac{V}{Z_N} &
0 \\[2ex]
0 & 0 & Z_\phi \lich_0+ m^2
\end{array}
\right)} \nn
&\qquad \small{
+\frac{1}{\tilde\a} \left(
\begin{array}{ccc}
\frac{9}{16} \lich_0\left(\lich_0 -\frac{\br}{3}\right)^2 &
 \frac{3\b}{16}\lich_0\left(\lich_0-\frac{\br}{3}\right) &0\\[2ex]
\frac{3\b}{16}\lich_0\left(\lich_0-\frac{\br}{3}\right) & \frac{\b^2}{16}\lich_0 & 0 \\[2ex]
0 & 0 & 0
\end{array}
\right),}
\end{align}
while the vertices $\mathcal V$ arising from $H_1^{\mu\nu}$, $H_2^{\mu\nu}$ and $V_1$ have the $\bar\phi$ dependence.

\section{Flow generators}
\label{app: sec: flow equation}

In this appendix, we explicitly show contributions from each mode in the metric fluctuation.
The regulator matrices $\mathcal R_k$ replace the Laplacians $z=\Delta_L$ in $\mathcal K$ to $P_k(z)=z+R_k(z)$.
In this work, we employ the Litim cutoff function $R_k(z)=(k^2-z)\theta(k^2-z)$.

Denoting the regulated kinetic terms by $\tilde{\mathcal K}$, the flow equation is expanded into polynomials of $\mathcal V$ as
\begin{align}
    \partial_t \Gamma_k &=\frac{1}{2}\Tr \left[ \left( \tilde{\mathcal K} + \mathcal V \right)^{-1} \partial_t \mathcal R_k\right]\nn
    &=\frac{1}{2}\Tr \left[ \tilde{\mathcal K}^{-1} \partial_t \mathcal R_k\right]
    -\frac{1}{2}\Tr \left[ \tilde{\mathcal K}^{-1}   \partial_t \mathcal R_k\tilde{\mathcal K}^{-1} \mathcal V \right]
    + \Tr \left[\tilde{\mathcal K}^{-1}   \partial_t \mathcal R_k\tilde{\mathcal K}^{-1} \mathcal V \tilde{\mathcal K}^{-1}\mathcal V \right] +\cdots.
    \label{app: expansion of flow equation}
\end{align}
Loop corrections to the cosmological constant and the Planck mass are involved in the first term in Eq.~\eqref{app: expansion of flow equation}, while those to the scalar potential $U(\rho)$ and the field renormalization factor $Z_\phi$ are obtained from the higher-order terms.
The Hessians of the spin-2 TT and the spin-1 transverse vector modes have no dependence on the scalar potential $U$, so those does not induce quantum corrections to scalar-field interactions,
while the spin-0 trace mode ($h$) has a $U$-dependence in $\mathcal K_{hh}$ and the mixing terms ($\mathcal V_{h\varphi}$, $\mathcal V_{\varphi h}$) from which the scalar potential receives quantum corrections from the trace mode.

\subsection{Heat kernel expansion}
\label{app: sec: heat kernel}
Before evaluating flow generators for each mode, we briefly summarize the heat kernel technique.
The flow generator in this work typically takes a form
\begin{align}
 \zeta =  \frac{1}{2}\Tr_{(i)} W[z=\Delta_i]= \frac{1}{2}\Tr_{(i)}\frac{\p_t  (aR_k(z))}{(a P_k(z) +b )^{p+1}}\bigg|_{(\Phi)},
 \label{app: typical flow generator}
\end{align}
where $a$, $b$ are scale-dependent constants and $i$ denotes the internal space of a field $\Phi$ on which the Laplacian acts, e.g. $i=$2TT, 1T, etc..
The heat kernel expansion gives
\begin{align}
  \zeta = \frac{1}{2(4\pi)^2}\int d^4x\sqrt{\bar g}\left[ b_0^{(i)}Q_2[W] + b_2^{(i)} Q_1[W] \bar R+\cdots \right].
\end{align}
Here $b_n^{(i)}$ are heat kernel coefficients and the threshold functions $Q_n$ are defined by
\begin{align}
&Q_n[W] =\frac{1}{\Gamma(n)} \int^\infty_0 dz\,z^{n-1}W[z],\qquad (\text{for $n\geq 1$}),\\[2ex]
&Q_{-n}[W] =(-1)^n\frac{\p^n W[z]}{\p z^n}\bigg|_{z=0} \qquad (\text{for $n\geq 0$}).
\end{align}
When employing the optimized cutoff \eqref{eq: optimized cutoff} for Eq.~\eqref{app: typical flow generator}, $Q_n[W]$ can be expressed in terms of the shorthand threshold function $\ell_p^{2n}$ introduced in Eq.~\eqref{eq: threshold function} as
\begin{align}
    Q_n[W]= a^{-p}\left( 1 -\frac{\eta_\Phi}{2(n+1)} \right) \left[2k^{2n-2p}\ell_p^{2n}(\tilde m_\Phi^2)\right],
\end{align}
where we have defined $\eta_\Phi=-\p_t a/a$ and $\tilde m_\Phi^2=b/(k^2a)$.
For the heat kernel coefficients of the Lichnerowicz Laplacians, see e.g. Ref.~\cite{Ohta:2016jvw}. Note that in a flat spacetime, the variable $z$ can be identified with loop momentum squared $q^2$ so  for $n\geq 1$
\begin{align}
    \frac{1}{2(4\pi)^2} Q_n[W] = \frac{1}{2}\int \frac{d^4q}{(2\pi)^4}\frac{q^{2n-4}}{\Gamma(n)} W[q^2].
\end{align}

\subsection{Transverse-traceless spin-2 mode}

The Hessian for the TT spin-2 mode is given in Eq~\eqref{eq: Hessian for TT mode}. 
Using the heat kernel expansion, one has
\begin{align}
\pi^\text{TT}
&= \frac{1}{2}\Tr \frac{\pa_t \mathcal R_k}{\G_k^{(2)}+ \mathcal R_k} \bigg|_{h^{TT}h^{TT}}\nn
&=\frac{1}{2}\Tr_\text{(2TT)}\left[ \frac{\pa_t R_k}{P_k -\frac{\bar R}{2}}\right]\nn
&= \frac{1}{(4\pi)^2}\int d^4x\sqrt{\bar g}
\left[
5\ell_0^4(0) +  k^2\left( -\frac{25}{6}\ell_0^2(0) + \frac{5}{2}\ell_0^4(0) \right)\bar R
\right].
\end{align}

\subsection{Spin-1 transverse vector modes}

In the system, there are three spin-1 transverse vector modes, i.e. $\xi_\mu$, $C^T_\mu$ ($\bar C^T_\mu$) and $J_\text{grav1}$ which have the same structure of the Hessian as in Eqs.~\eqref{ghostkin}, \eqref{eq: Hessian for xi} and \eqref{eq: Jaconians arising from decomposition}.
We denote these contributions by $\eta^{(1)}$ which is evaluated by the heat kernel technique:
\begin{align}
\eta^{(1)}&=\frac{1}{2} \Tr\frac{\partial_t  \mathcal  R_k}{\Gamma_k^{(2)}+  \mathcal R_k}\bigg|_{\xi\xi}
- \Tr\frac{\partial_t  \mathcal R_k}{\Gamma_k^{(2)}+  \mathcal R_k}\bigg|_{\bar C^T C^T}
- \frac{1}{2}\Tr\frac{\partial_t  \mathcal R_k}{\Gamma_k^{(2)}+ \mathcal R_k}\bigg|_{J_\text{grav1}}\nn
&= -\frac{1}{2}\Tr_\text{(1T)} \left[\frac{\partial_t R_k}{P_k - \frac{\bar R}{2}}\right]\nn
&= -\frac{1}{(4\pi)^2}\int d^4x\sqrt{\bar g}
\left[
3\ell_0^4(0) + k^2 \left( -\frac{1}{2}\ell_0^2(0) + \frac{3}{2}\ell_0^4(0) \right)\bar R
\right].
\end{align}

\subsection{Spin-0 scalar modes}

We have $(\sigma, h, \phi)$, $C^L$ ($\bar C^L$) and $J_\text{grav0}$ as spin-0 scalar modes. Among them, two fields should be physical.  One of physical modes is $\phi$, while another is a linear combination of $\sigma$ and $h$. Other modes are unphysical. Denoting  $\pi^\text{(0)}$ and $\eta^\text{(0)}$ physical and unphysical modes, respectively, their flow generators are given by
\begin{align}
\pi^\text{(0)}+ \eta^\text{(0)}
&=\frac{1}{2} \Tr\frac{\partial_t  \mathcal R_k}{\Gamma_k^{(2)}+ \mathcal  R_k}\bigg|_\text{scalar}
- \Tr\frac{\partial_t  \mathcal R_k}{\Gamma_k^{(2)}+ \mathcal R_k}\bigg|_{\bar C^L C^L}
- \frac{1}{2}\Tr\frac{\partial_t  \mathcal R_k}{\Gamma_k^{(2)}+  \mathcal R_k}\bigg|_{J_\text{grav0}}.
\label{eq: flow generator from scalar modes}
\end{align}
The last two terms can be simplified to be
\begin{align}
&- \Tr\frac{\partial_t  \mathcal R_k}{\Gamma_k^{(2)}+ \mathcal R_k}\bigg|_{\bar C^L C^L}
- \frac{1}{2}\Tr\frac{\partial_t  \mathcal R_k}{\Gamma_k^{(2)}+  \mathcal R_k}\bigg|_{J_\text{grav0}}\nn
&\qquad
=- \Tr_{(0)} \left[\frac{\partial_t R_k}{P_k - \frac{{\bar R}}{3-\beta}} \right]
-\frac{1}{2}\Tr_{(0)} \left[\frac{\partial_t R_k}{P_k - \frac{{\bar R}}{3}}\right]
-\frac{1}{2}\Tr_{(0)} \left[\frac{\partial_t R_k}{P_k} \right]\nn
&\qquad
= -\frac{1}{(4\pi)^2}\int d^4x\sqrt{\bar g}
\left[
4\ell_0^4(0) +  k^2\left( \frac{2}{3}\ell_0^2(0) + \frac{9-\beta}{3(3-\beta)}\ell_0^4(0) \right)\bar R
\right].
\end{align}
The first term on the rhs in Eq.~\eqref{eq: flow generator from scalar modes} is evaluated as follows: We first calculate the inverse matrix of
the Hessian \eqref{eq: Hessian for scalar fields} with the regulator matrix $\mathcal R_k$ which replaces the Laplacians
in the Hessian by $P_k$. Then, we evaluate the trace for the product of the inverse matrix of the Hessian and the regulator matrix differentiated by the scale, $\partial_t \mathcal R_k$.
In this way, one obtains
\begin{align}
    &\frac{1}{2} \Tr\frac{\partial_t  \mathcal R_k}{\Gamma_k^{(2)}+ \mathcal  R_k}\bigg|_\text{scalar}
    =\frac{1}{(4\pi)^2}\int d^4x\sqrt{\bar g}\left[ k^4S_2(\tilde\rho)  +k^2\left( 
     S_1(\tilde \rho)
    +\frac{1}{6}S_2(\tilde\rho) \right)\bigg|_{\tilde\rho=0} \bar R \right],
\end{align}
where
\begin{align}
    S_2(\tilde\rho)&=3\ell_0^4(0)
    +\ell_0^4(-\tilde M_s^2)\nn
    &\qquad + 2\left(1-\tfrac{\eta_\phi}{6}\right)\left[1 -2 \frac{4(3-\tilde\alpha)}{(3-\beta)^2\tilde M_P^2}\tilde U(\tilde\rho) \right]\ell_0^4(-\tilde M_s^2(\tilde \rho)) \ell_0^4(\tilde M_H^2(\tilde\rho)),\\
    S_1(\tilde\rho=0)&= \frac{6-\tilde\alpha}{6(3-\tilde\alpha)}
    -\left[ \frac{3-\tilde\alpha}{(3-\beta)^2} - \frac{1}{3-\beta} +\frac{1}{3-\tilde\alpha} \right]\ell_1^2(-\tilde m_s^2).
\end{align}

\subsection{Anomalous dimension of the scalar field}
\label{app: sec: anomalous dimension of scalar field}

Let us derive the anomalous dimension of the scalar field, $\eta_\phi$ induced by the metric fluctuations.
The full result is shown in Eq.~\eqref{anomalous dimension etaphi}.

We first note that we can consider a flat spacetime background $\bar g_{\mu\nu}=\delta_{\mu\nu}$ to obtain $\eta_\phi$, so the Fourier transformation for fields can be employed, i.e.
\begin{align}
    \frac{Z_\phi}{2}\int d^4x\,\delta^{\mu\nu}(\partial_\mu \phi)(\partial_\nu\phi)
    =\frac{Z_\phi}{2}\int \frac{d^4q}{(2\pi)^4} q^2 \phi(q) \phi(-q).
\end{align}
The flow equation for the renomalization factor of the scalar field is obtained by evaluating
\begin{align}
\partial_tZ_\phi=\frac{1}{\Omega}\frac{d}{dp^2}\frac{\delta}{\delta \bar\phi(p)}\frac{\delta}{\delta \bar\phi(-p)} \partial_t \Gamma_k\Big|_{p^2=0},
\end{align}
where $\Omega=\int d^4x=(2\pi)^4\delta^4(0)$ is a four-dimensional spacetime volume.
Diagrammatically, there are the following contributions: From Eq.~\eqref{app: expansion of flow equation},
\begin{align}
    \eta_\phi&=
  -\frac{1}{Z_\phi}\frac{1}{\Omega}\frac{d}{dp^2}\frac{\delta}{\delta \bar\phi(p)}\frac{\delta}{\delta \bar\phi(-p)} \left.\left(-\frac{1}{2}\Tr \left[ \tilde{\mathcal K}^{-1}   \partial_t \mathcal R_k\tilde{\mathcal K}^{-1} \mathcal V \right]
    + \Tr \left[\tilde{\mathcal K}^{-1}   \partial_t \mathcal R_k\tilde{\mathcal K}^{-1} \mathcal V \tilde{\mathcal K}^{-1} \mathcal V\right] \right)\right|_{p^2=0} \nn
    &=
    -\frac{1}{Z_\phi}\frac{1}{\Omega}\frac{d}{dp^2}
\left.\left(-\frac{1}{2}\vcenter{\hbox{\includegraphics[bb=0 0 131 87, width=22mm]{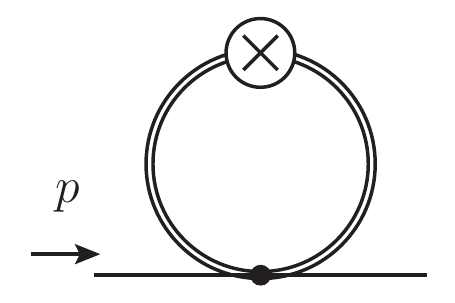}}}
+\vcenter{\hbox{\includegraphics[bb=0 0 164 72, width=24mm]{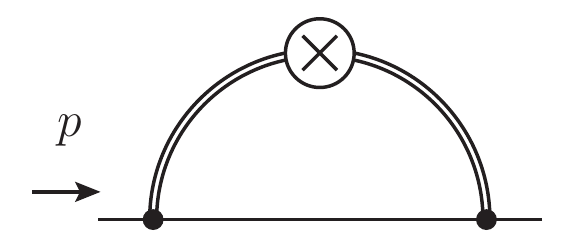}}}
+ \vcenter{\hbox{\includegraphics[bb=0 0 164 70, width=24mm]{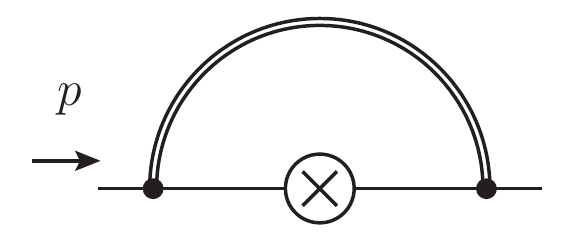}}} \right)\right|_{p^2=0},
\label{app: anomalous dimension of phi}
\end{align}
where solid and double-solid lines represent a scalar field and metric fluctuation fields $h_{\mu\nu}$, respectively, and the cross-circle denotes the cutoff insertion in the propagator.
For the first diagram (tadpole diagram), a vertex $H_2^{\mu\nu}$ given in Eq.~\eqref{H2 explicit} contributes, whereas the second and third diagrams (sunset diagrams) have two vertices of $H_1^{\mu\nu}$.
Below we calculate these contributions explicitly. 

\subsubsection{Tadpole diagram}
Before computing contributions from the tadpole diagram, we consider the trace of $H_2^{\mu\nu}$, namely
\begin{align}
\delta_{\mu\nu}H_2^{\mu\nu} =\frac{1}{2}h_{\mu\nu} h^{\mu\nu}.
\end{align}
Hence, due to the functional trace in the flow equation, the first and third terms in Eq.~\eqref{H2 explicit} cancel each other, and thus only the second term in Eq.~\eqref{H2 explicit} contributes.
Moreover, as shown in Eq.~\eqref{app: squared metric fluctuation}, the squared metric fluctuation field has no mixing term between different modes.

First, we evaluate the contribution from a TT-mode loop which is denoted by a double wiggly line below.
\begin{align}
\eta_\phi\Big|_\text{tadpole}^\text{2TT}&=-\frac{1}{Z_\phi}\frac{1}{\Omega}\frac{d}{dp^2}\left.\left(
-\frac{1}{2}\vcenter{\hbox{\includegraphics[bb=0 0 145 97, width=22mm]{tadpole_tt}}}
\right)\right|_{p^2=0}\nn
&=\frac{1}{Z_\phi}\frac{d}{dp^2}\frac{1}{\Omega}\frac{1}{2}\Tr
\left[ \tilde{\mathcal K}^{-1} \partial_t {\mathcal R}_k \tilde{\mathcal K}^{-1} ({\mathcal V})^{\mu\nu}{}_{\rho\sigma} (P_{TT})^{\rho\sigma}{}_{\mu\nu}
\right]_{h^{TT}h^{TT}}
\nn
&=\frac{1}{Z_\phi}\frac{d}{dp^2}\frac{1}{2}\Tr
\left[
\frac{\partial_t R_k(q)}{P_k(q)^2}\left(\frac{2Z_\phi}{Z_N} \delta^\nu_{\sigma} (p^\mu p_\rho) (P_{TT}(q))^{\rho\sigma}{}_{\mu\nu}  \right)
\right]\nn
&=\frac{5}{2Z_N}
\frac{1}{2}\int\frac{d^4q}{(2\pi)^4}\frac{\partial_t R_k(q)}{P_k(q)^2}
=\frac{5}{(4\pi)^2}\frac{1}{\tilde M_P^2}\ell_1^4(0),
\label{app: tadpole of TT mode}
\end{align}
where $(P_{TT}(q))^{\rho\sigma}{}_{\mu\nu}$ is defined in Eq.~\eqref{app: TT projector}.

Second, the loop effect of the transverse spin-1 mode is calculated. Denoting it by a single wiggly line, we obtain 
\begin{align}
\eta_\phi\Big|_\text{tadpole}^\text{1T}&=
-\frac{1}{Z_\phi}\frac{1}{\Omega}\frac{d}{dp^2}\left.\left(
-\frac{1}{2}\vcenter{\hbox{\includegraphics[bb=0 0 145 97, width=22mm]{tadpole_spin1}}}
\right)\right|_{p^2=0} \nn
&=\frac{1}{Z_\phi}\frac{d}{dp^2}\frac{1}{\Omega}\frac{1}{2}\Tr
\left[ \tilde{\mathcal K}^{-1} \partial_t {\mathcal R}_k \tilde{\mathcal K}^{-1} ({\mathcal V})^{\mu}{}_{\nu}  (P_1)^{\nu}{}_{\mu}
\right]_{\xi \xi}
\nn
&=\frac{1}{Z_\phi}\frac{d}{dp^2}\frac{1}{2}\Tr
\left[
\frac{2P_k\partial_t R_k(q)}{P_k(q)^4}\left(\frac{\tilde\alpha Z_\phi}{Z_N}q^2\left\{ \frac{1}{4}\delta^{\mu\nu}\delta^{\alpha\beta}+\delta^{\alpha\mu}\delta^{\beta\nu} \right\} (P_1(q))_{\alpha\beta}p_\mu p_\nu  \right)
\right]\nn
&=\frac{3\tilde\alpha}{2Z_N}
\int\frac{d^4q}{(2\pi)^4}q^2\frac{\partial_t R_k(q)}{P_k(q)^3}
=\frac{12   \tilde\alpha}{(4\pi)^2}\frac{1}{\tilde M_P^2}\ell_2^6(0).
\label{app: tadpole of spin 1 mode}
\end{align}
Note that we have redefined the transverse vector mode as $\xi_\mu \to \tilde\alpha^{1/2}\xi_\mu$.

Finally, let us calculate contributions from the scalar modes in the metric fluctuations for which we need the $2\times 2$ vertex matrix,
\begin{align}
    \pmat{
{\mathcal V}_{\sigma\sigma} & {\mathcal V}_{\sigma h} \\[2ex]
{\mathcal V}_{h\sigma} & {\mathcal V}_{hh}
}.
\end{align}
Their propagators are given as a $2\times 2$ matrix, i.e. the regulated propagator matrix reads
\begin{align}
\tilde{\mathcal K}^{-1}\Big|_{2\times 2}  = 
\frac{16}{P_k - k^2 \tilde m_s^2}
\begin{pmatrix}
 -\frac{1}{3 P_k^2}\frac{(\beta ^2-3 \tilde\alpha ) P_k +4 \tilde\alpha  k^2 v_0}{(\beta -3)^2P_k}
 && \frac{(\beta -\tilde\alpha)}{(\beta -3)^2P_k} \\[2ex]
 \frac{(\beta -\tilde\alpha )}{(\beta -3)^2P_k} && \frac{(\tilde\alpha -3)}{(\beta -3)^2}
\end{pmatrix},
\label{app: 2times2 propagator}
\end{align}
with $\tilde m_s^2=\frac{4(3-\tilde\alpha )}{(3-\beta)^2}v_0$ and the regulator matrix
\begin{align}
\mathcal R_k\Big|_{2\times 2}=\frac{1}{16}
\begin{pmatrix}
 \frac{3 (3-\tilde\alpha)}{\tilde\alpha}\left(P_k^3-q^6\right) & \frac{3 (\beta-\tilde\alpha)}{\alpha } \left(P_k^2-q^4\right) \\[2ex]
  \frac{3 (\beta-\tilde\alpha)}{\tilde\alpha } \left(P_k^2-q^4\right)& 
  \frac{\beta^2- 3 \tilde\alpha}{\tilde \alpha } \left(P_k-q^2\right)
\end{pmatrix}.
\label{app: 2times 2 regulator matrix}
\end{align}
As mentioned above, the traced $H_2^{\mu\nu}$ contains no mixing vertex between $\sigma$ and $h$, so we define the vertex matrix of scalar modes as the following diagonal form
\begin{align}
\pmat{
{\mathcal V}_{\sigma\sigma} & {\mathcal V}_{\sigma h} \\[2ex]
{\mathcal V}_{h\sigma} & {\mathcal V}_{hh} 
}
\to
{\mathcal V}\Big|_{2\times 2} =\frac{Z_\phi}{Z_N}\Omega \frac{p^2}{16}
\begin{pmatrix}
3q^4 & 0 \\[1ex]
0 & 1
\end{pmatrix},
\end{align}
where we already performed the symmetrization for the loop momenta; $q^\mu q^\nu \to \frac{q^2}{4}\delta^{\mu\nu}$ and $p_\mu$ are external momenta.

We denote the propagator of scalar modes in the metric fluctuations by a dashed line and evaluate contributions from their tadpole diagram as follows:
\begin{align}
\eta_\phi\Big|_\text{tadpole}^\text{scalar}&=
-\frac{1}{Z_\phi}\frac{1}{\Omega}\frac{d}{dp^2}\left.\left(
-\frac{1}{2}\vcenter{\hbox{\includegraphics[bb=0 0 145 91, width=22mm]{tadpole_spin0}}}
\right)\right|_{p^2=0}\nn
&=\frac{1}{Z_\phi}\frac{1}{\Omega}\frac{d}{dp^2}\frac{1}{2}\Tr
\left[
\tilde{\mathcal K}^{-1} 
\partial_t {\mathcal R}_k 
\tilde{\mathcal K}^{-1}{\mathcal V}
\right]_{2\times 2}\nn
&=-\frac{1}{(4\pi)^2}\frac{4(3-\tilde\alpha)}{(3-\beta)^2\tilde M_P^2}\Bigg[
\frac{1}{4}\ell_1^2(-\tilde m_s^2) - \frac{9\tilde\alpha(3-\beta)^2}{(3-\tilde\alpha)^2}\ell_0^8(0)\nn
&\qquad
+\frac{9(\tilde\alpha-\beta)^2}{(3-\tilde\alpha)^2}\left( \ell_1^6(-\tilde m_s^2) + 2 \ell_0^8(-\tilde m_s^2) \right)
\Bigg].
\label{app: tadpole of spin 0 mode}
\end{align}

To summarize, the total contributions from the tadpole diagrams are the sum of Eqs.~\eqref{app: tadpole of TT mode}, \eqref{app: tadpole of spin 1 mode} and \eqref{app: tadpole of spin 0 mode}.

Here we briefly comment on the case of the linear parametrization for which 
the vertex with two-metric fluctuations reads 
\begin{align}
H_2^{\mu\nu}= \left(-\frac{1}{4}h^{\alpha\beta}h_{\alpha\beta} +\frac{1}{8}h^2 \right) \bar g^{\mu\nu} + h^\mu{}_\lambda h^{\nu\lambda}  - \frac{1}{2}h h^{\mu\nu}.
\end{align}
Taking the trace for this, we find that $\bar g_{\mu\nu}H_2^{\mu\nu}=0$, thus there are no contributions from the tadpole diagrams in the linear parametrization.

\subsubsection{Sunset diagrams}

We compute the anomalous dimension of the scalar field from the sunset diagrams which are given by last two terms in Eq.~\eqref{app: anomalous dimension of phi}.
We note that although the three-point vertex $\bar\phi$-$h$-$\varphi$ arises from the scalar mass term $m^2\sqrt{g}\phi^2$, we do not take it into account by assuming that scalar interactions have only the Gaussian fixed point.
Thus, we consider contributions from only $H_1^{\mu\nu}$.

We first recognize that the sunset diagrams with the TT mode propagator does not contribute because within the flow generator, loop momenta $q^\mu$ included in the vertex $\mathcal V_{h^{TT}\varphi}$ contract with the TT projector \eqref{app: TT projector} and give $q^\mu (P_{TT}(q))^{\rho\sigma}{}_{\mu\nu}=0$.

We next consider the anomalous dimension $\eta_\phi$ induced by the interactions $\mathcal V_{\xi\varphi}$ and $\mathcal V_{\varphi\xi}$.
The flow equation contains the following algebraic computation: 
\begin{align}
\Tr \Big[(iq^\rho \delta_\alpha^\sigma) q_\rho p_\sigma (iq^\mu \delta^\nu_{\beta}) q_\mu p_\nu (P_1(q))^{\alpha\beta} \Big]
=- \frac{3}{4}q^2 p^2.
\end{align}
Using this, we obtain
\begin{align}
\eta_\phi\Big|_\text{sunset}^\text{1T}&=
-\frac{1}{Z_\phi}\frac{1}{\Omega}\frac{d}{dp^2}\left.\left(
\vcenter{\hbox{\includegraphics[bb=0 0 137 71, width=23mm]{sunset_spin12}}}
+\vcenter{\hbox{\includegraphics[bb=0 0 137 72, width=23mm]{sunset_spin11}}}
\right)\right|_{p^2=0}\nn
&=-\frac{1}{Z_\phi}\frac{1}{\Omega}\frac{d}{dp^2}\Tr \left[\tilde{\mathcal K}^{-1}   \partial_t \mathcal R_k\tilde{\mathcal K}^{-1} \mathcal V \tilde{\mathcal K}^{-1}\mathcal V \right]_{\xi \varphi}\nn
&=\frac{3}{4}\frac{{\tilde \alpha}}{Z_N}\Tr \Bigg[ \frac{Z_\phi}{Z_\phi P_k(q) + m^2}\frac{q^2\partial_t R_k(q)}{P_k(q)^2}
\Bigg]
+\frac{3}{4}\frac{{\tilde \alpha}}{Z_N}\Tr \Bigg[ \frac{Z_\phi q^2\partial_t (Z_\phi R_k(q))}{(Z_\phi P_k(q) +m^2)^2}\frac{1}{P_k(q)}
\Bigg] \nn
&={\tilde \alpha}\frac{1}{Z_N} \frac{3}{4}\int \frac{d^4q}{(2\pi)^4}\frac{q^2\partial_t R_k}{P_k^2} \frac{1}{P_k+ \tilde m_H^2 k^2} 
+{\tilde \alpha}\frac{1}{Z_N} \frac{3}{4}\int \frac{d^4q}{(2\pi)^4} \frac{q^2\partial_t (Z_\phi R_k)/Z_\phi}{(P_k+ \tilde m_H^2 k^2)^2}\frac{1}{P_k}\nn
&=\tilde \alpha \frac{6}{(4\pi)^2}\frac{1}{\tilde M_P^2}\left[
\ell_1^6(0)\ell_0^2(\tilde m_H^2)
 + \left(1 - \tfrac{\eta_\phi}{6} \right)\ell_1^6(\tilde m_H^2)\ell_0^2(0)
 \right],
 \label{eq:sunset1}
\end{align}
where we have used $q^\mu (P_1(q))_{\mu\nu}=0$ and $p_\beta (P_1)^{\beta \alpha}p_{\alpha}=3p^2/4$.

Next, we evaluate loop effects of spin-0 modes for which the following $3\times 3$ vertex matrix is needed:
\begin{align}
\pmat{
0 & 0 & {\mathcal V}_{\sigma\varphi}\\
0 & 0 & {\mathcal V}_{h\varphi}\\
{\mathcal V}_{\varphi\sigma} & {\mathcal V}_{\varphi h} & 0
}.
\end{align}
The flow equation for the two-point function of $\bar\phi$ is given by
\begin{align}
&\frac{\delta^2(\partial_t \Gamma_k)}{\delta \bar\phi(p)\delta\bar\phi(-p)}\nn
&=
\frac{\delta^2}{\delta \bar\phi(p)\delta\bar\phi(-p)}\Tr \left[
\pmat{
0 & 0 & {\mathcal V}_{\sigma\varphi}\\
0 & 0 & {\mathcal V}_{h\varphi}\\
{\mathcal V}_{\varphi\sigma} & {\mathcal V}_{\varphi h} & 0
}\right. 
\pmat{
\tilde{\mathcal K}_{\sigma\sigma} &\tilde{\mathcal K}_{\sigma h} & 0\\
\tilde{\mathcal K}_{\sigma h} & \tilde{\mathcal K}_{hh} & 0\\
0 & 0 &\tilde{\mathcal K}_{\varphi\varphi}
}^{-1}
\pmat{
\partial_t R_{\sigma\sigma} &\partial_t R_{\sigma h} & 0\\
\partial_t R_{\sigma h} & \partial_t R_{hh} & 0\\
0 & 0 & \partial_t R_{\varphi\varphi}
}
\nn
&\qquad\qquad
\left.
\times
\pmat{
\tilde{\mathcal K}_{\sigma\sigma} &\tilde{\mathcal K}_{\sigma h} & 0\\
\tilde{\mathcal K}_{\sigma h} & \tilde{\mathcal K}_{hh} & 0\\
0 & 0 &\tilde{\mathcal K}_{\varphi\varphi}
}^{-1}
\pmat{
0 & 0 & {\mathcal V}_{\sigma\varphi}\\
0 & 0 & {\mathcal V}_{h\varphi}\\
{\mathcal V}_{\varphi\sigma} & {\mathcal V}_{\varphi h} & 0
}
\pmat{
\tilde{\mathcal K}_{\sigma\sigma} & \tilde{\mathcal K}_{\sigma h} & 0\\
\tilde{\mathcal K}_{\sigma h} & \tilde{\mathcal K}_{hh} & 0\\
0 & 0 &\tilde{\mathcal K}_{\varphi\varphi}
}^{-1}
 \right] \nn
&=\frac{\delta^2}{\delta \bar\phi(p)\delta\bar\phi(-p)}\Tr
\Bigg[
\frac{1}{\tilde{\mathcal K}_{\varphi\varphi}}
\pmat{
{\mathcal V}_{\sigma \varphi}{\mathcal V}_{\varphi\sigma} & {\mathcal V}_{\sigma \varphi}{\mathcal V}_{\varphi h} \\
{\mathcal V}_{h \varphi}{\mathcal V}_{\varphi \sigma}  & {\mathcal V}_{h \varphi}{\mathcal V}_{\varphi h}
}
\left[\tilde{\mathcal K}^{-1} 
\partial_t {\mathcal R}_k 
\tilde{\mathcal K}^{-1}
\right]_{2\times 2}
\Bigg] 
\nn
&\qquad
+\frac{\delta^2}{\delta \bar\phi(p)\delta\bar\phi(-p)}\Tr \Bigg[
\frac{\partial_t R_{\varphi\varphi}}{\tilde{\mathcal K}_{\varphi\varphi}^2}
\pmat{
{\mathcal V}_{\sigma\varphi}{\mathcal V}_{\varphi\sigma}
& {\mathcal V}_{\sigma\varphi}{\mathcal V}_{\varphi h} \\
{\mathcal V}_{h \varphi}{\mathcal V}_{\varphi \sigma}  & 
{\mathcal V}_{h\varphi}{\mathcal V}_{\varphi h}
}
\tilde{\mathcal K}^{-1}\Big|_{2\times 2}
 \Bigg]\nn
&=
\vcenter{\hbox{\includegraphics[bb=0 0 137 71, width=29mm]{sunset_spin02}}}
+\vcenter{\hbox{\includegraphics[bb=0 0 137 67, width=29mm]{sunset_spin01}}}.
\label{app: second derivative of Gamma in scalar modes}
\end{align}
Here, 
$\tilde{\mathcal K}^{-1}|_{2\times 2}$ and $\mathcal R_k|_{2\times 2}$ are defined in Eqs.~\eqref{app: 2times2 propagator} and \eqref{app: 2times 2 regulator matrix}, respectively, and the vertex matrix is reduced to the $2\times 2$ matrix
\begin{align}
&\frac{\delta^2}{\delta \bar\phi(p)\delta\bar\phi(-p)}\pmat{
{\mathcal V}_{\sigma \varphi}{\mathcal V}_{\varphi\sigma} & {\mathcal V}_{\sigma \varphi}{\mathcal V}_{\varphi h} \\
{\mathcal V}_{\sigma \varphi}{\mathcal V}_{\varphi h}  & {\mathcal V}_{h \varphi}{\mathcal V}_{\varphi h}
}\nn
&=\Omega\frac{Z_\phi^2}{M_P^2}
\pmat{
2\left( q^\mu q^\nu -\frac{\delta^{\mu\nu}}{4}q^2 \right) q_\mu p_\nu 
\times 2\left( q^\rho q^\sigma -\frac{\delta^{\rho\sigma}}{4}q^2 \right) q_\rho p_\sigma && 2\left( q^\mu q^\nu -\frac{\delta^{\mu\nu}}{4}q^2\right) q_\mu p_\nu  \times \frac{1}{2}\delta^{\rho\sigma} q_\rho p_\sigma \\[2ex]
2\left( q^\mu q^\nu -\frac{\delta^{\mu\nu}}{4}q^2\right) q_\mu p_\nu  \times\frac{1}{2}\delta^{\rho\sigma} q_\rho p_\sigma && \frac{1}{2}\delta^{\mu\nu} q_\mu p_\nu \times \frac{1}{2}\delta^{\rho\sigma} q_\rho p_\sigma
} \nn[1ex]
&\to 
\frac{Z_\phi^2}{M_P^2}\Omega\frac{p^2}{16}
\pmat{
9q^6  & 3q^4  \\[1ex]
3q^4  & q^2
},
\end{align}
where in the last step, we symmetrized loop momenta by $q^\mu q^\nu \to \frac{q^2}{4}\delta^{\mu\nu}$.

Continuing the evaluation of Eq.~\eqref{app: second derivative of Gamma in scalar modes}, the scalar modes in the metric fluctuations gives
\begin{align}
\eta_\phi\Big|_\text{sunset}^\text{scalar}&=-\frac{1}{Z_\phi}\frac{1}{\Omega}\frac{d}{dp^2}\left.\left(
\vcenter{\hbox{\includegraphics[bb=0 0 137 71, width=23mm]{sunset_spin02}}}
+\vcenter{\hbox{\includegraphics[bb=0 0 137 67, width=23mm]{sunset_spin01}}}
\right)\right|_{p^2=0}\nn
&=-\frac{1}{Z_\phi}\frac{d}{dp^2}\Tr
\Bigg[
\frac{1}{Z_\phi(P_k + k^2 \tilde m_H^2)}
\frac{Z_\phi^2}{Z_N}\frac{p^2}{16}
\pmat{
9q^6  & 3q^4  \\[1ex]
3q^4  & q^2
}
\left[\tilde{\mathcal K}^{-1} 
\partial_t {\mathcal R}_k 
\tilde{\mathcal K}^{-1}
\right]_{2\times 2}
\nn
&\qquad
-\frac{1}{Z_\phi}\frac{d}{dp^2}\Tr \Bigg[
\frac{\partial_t (Z_\phi R_k)}{Z_\phi^2(P_k+ k^2 \tilde m_H^2)^2}
\frac{Z_\phi^2}{Z_N}\frac{p^2}{16}
\pmat{
9q^6  & 3q^4  \\[1ex]
3q^4  & q^2
}
\tilde{\mathcal K}^{-1} \Big|_{2\times 2}
 \Bigg]\nn
&= \frac{2}{(4\pi)^2}\frac{4(3-\tilde\alpha)}{(3-\beta)^2\tilde M_P^2}\Bigg[ \ell_0^2(\tilde m_H^2)\Bigg(
\ell_1^6(-\tilde m_s^2)
+ \frac{18 ({\tilde \alpha} -\beta )}{(3-\tilde\alpha)} \left(  \ell_1^8(-\tilde m_s^2)+ \ell_0^8(-\tilde m_s^2) \right) \nn
&\quad
+ \frac{108({\tilde \alpha} -\beta )^2}{(3-{\tilde \alpha} )^2} \left( \ell_1^{10}(-\tilde m_s^2) +2 \ell_0^{10}(-\tilde m_s^2)  \right) 
- \frac{108{\tilde \alpha} (3-\beta)^2}{(3-{\tilde \alpha})^2}\ell_0^{10}(0)
 \Bigg) \nn
&\qquad
+\ell_1^2(\tilde m_H^2) \Bigg(
 \left( 1 -\tfrac{\eta_\phi}{8} \right)\ell_0^6(-\tilde m_s^2)
+  \left( 1-\tfrac{\eta_\phi}{10} \right) \frac{18 ({\tilde \alpha} -\beta )}{(3-\tilde\alpha )}\ell_0^8(-\tilde m_s^2) \nn
&\quad
+ \left( 1 -\tfrac{\eta_\phi}{12}  \right) \left(\frac{108({\tilde \alpha} -\beta )^2}{(3-{\tilde \alpha})^2}\ell_0^{10}(-\tilde m_s^2) - \frac{36 {\tilde \alpha} (3-\beta)^2 }{(3-{\tilde \alpha} )^2}\ell_0^{10}(0) \right)
\Bigg)
\Bigg].
\label{eq:sunset2}
\end{align}

The total contributions from sunset diagrams to the anomalous dimension of the scalar field is the sum of Eqs.~\eqref{eq:sunset1}
and \eqref{eq:sunset2}.


\printbibliography
\end{document}